\documentclass[a4paper, 11pt]{article}
\usepackage{epsfig,amsmath,amsfonts,hyperref,wasysym,tikz}

\newcommand{\be}{\begin{equation}}
\newcommand{\ee}{\end{equation}}

\newcommand{\triplet}[3]{\left(\begin{array}{c}#1\\#2\\#3\end{array}\right)}

\newcommand{\half}{\frac12}

\newcommand{\Urm}{\mathrm{U}}
\newcommand{\su}{\mathrm{su}}

\newcommand{\SU}{\mathrm{SU}}
\newcommand{\SO}{\mathrm{SO}}
\newcommand{\SL}{\mathrm{SL}}

\newcommand{\Tr}{\mathrm{Tr}}
\newcommand{\Acal}{\mathcal{A}}

\newcommand{\Wcal}{\mathcal{W}}

\newcommand{\Zset}{\mathbb{Z}}
\newcommand{\Ncal}{\mathcal{N}}

\newcommand{\Vcal}{\mathcal{V}}
\newcommand{\AdS}{\mathrm{AdS}}
\newcommand{\Srm}{\mathrm{S}}
\newcommand{\Cset}{{\,\,{{{^{_{\pmb{\mid}}}}\kern-.47em{\mathrm C}}}}}

\newcommand{\ket}[1]{\left|#1\right.\rangle}
\newcommand{\qb}{{\bar{q}}}
\newcommand{\hb}{{\bar{h}}}
\newcommand{\zb}{{\bar{z}}}

\newcommand{\ra}{\rightarrow}

\newcommand{\bb}{\bar{b}}
\newcommand{\fb}{\bar{f}}
\newcommand{\astv}{\!\mathrel{\raisebox{.8pt}{{\scriptsize \hexstar}}}\!}

\newcommand{\darkblue}{blue!70!black!75!white}
\newcommand{\darkred}{red!70!black!60!orange}
\newcommand{\darkgreen}{green!50!black}
\newcommand{\graphite}{black!71!white}
\newcommand{\cdb}{\color{\darkblue}}

\newcommand{\cdr}{\color{\darkred}}

\newcommand{\trr}{\triangleright}
\newcommand{\id}{\mathrm{id}}

\newcommand{\comment}[1]{}

\newcommand{\smin}{\raisebox{1.2pt}{\tiny $-$}}
\newcommand{\splus}{\raisebox{1.2pt}{\tiny $+$}}

\begin{document}

\numberwithin{equation}{section}

\mbox{}

\vspace{40pt}

\begin{center}

 {\Large \bf Marginal deformations and quasi-Hopf algebras}\\
\vspace{43pt}

{\large {\mbox{{\bf Hector Dlamini$\,{}^a$} \hspace{.2cm} and \hspace{.2cm} {\bf Konstantinos Zoubos$\,{}^b$}}}}%

\vspace{.5cm}

Department of Physics, University of Pretoria\\
Private Bag X20, Hatfield 0028, South Africa

\mbox{}

and

\mbox{}

National Institute for Theoretical Physics (NITheP) \\
Gauteng, South Africa

\vspace{40pt}

{\Large \bf Abstract}

\end{center}

\vspace{.3cm}

\large

\noindent We establish the existence of a quasi-Hopf algebraic structure underlying the Leigh-Strassler
$\Ncal=1$ superconformal marginal deformations of the $\Ncal=4$ Super-Yang-Mills theory. The scalar-sector
$R$-matrix of these theories, which is related to their one-loop spin chain Hamiltonian, does not generically
satisfy the Quantum Yang-Baxter Equation. By constructing a Drinfeld twist which relates this $R$-matrix to
that of the $\Ncal=4$ SYM theory, but also produces a non-trivial co-associator, we show that the generic
Leigh-Strassler $R$-matrix satisfies the quasi-Hopf version of the QYBE. We also use the twist to define
a suitable star product which directly relates the $\Ncal=4$ SYM superpotential to that of the marginally
deformed gauge theories. We expect our results to be relevant to studies of integrability (and its breaking)
in these theories, as well as to provide useful input for supergravity solution-generating techniques.

\normalsize

\noindent

\vspace{3.6cm}
\noindent\rule{4.5cm}{0.4pt}

\noindent $^a$ sickmech@gmail.com 

\noindent $^b$ kzoubos@up.ac.za

\vspace{0.5cm}

\setcounter{page}{0}
\thispagestyle{empty}
\newpage

\tableofcontents

\section{Introduction}

The superconformal $\Ncal=4$ Super-Yang-Mills (SYM) theory plays a primary role in modern theoretical
physics. Among its many special features one can highlight its (still conjectural) exact duality with a string
theory via the AdS/CFT correspondence \cite{Maldacena:1997re} and its all-loop planar integrability, which arises by
mapping to a spin chain model \cite{Minahan:2002ve}. One may ask to what extent these features persist as we
deform away from the maximally supersymmetric theory towards models with less symmetry. In this work we will
focus on  marginal deformations of $\Ncal=4$ SYM, which preserve the conformal invariance of the original
theory but reduce the amount of supersymmetry. 

It has been known for some time  that $\Ncal=4$ SYM admits a class of $\Ncal=1$ exactly
marginal deformations, with a non-perturbative argument
provided by Leigh and Strassler \cite{Leigh:1995ep}. In the formalism of $\Ncal=1$ superspace, the
superpotential of these more general theories takes the form
\be \label{LSW}
\Wcal_{LS}=
\kappa~{\mathrm {Tr}}\left(\Phi^1~[\Phi^2,\Phi^3]_{q}+ \frac h3\left((\Phi^1)^3+(\Phi^2)^3+(\Phi^3)^3\right)\right)\;,
\ee
with $[X,Y]_q=XY-qYX$. The parameters $\kappa, q$ and $h$ are generically complex, so, together
with the complexified gauge coupling $\tau=\!\theta/2\pi\!+4\pi i/g^2$, the theory depends on four complex parameters\footnote{However,
we will set the $\theta$-angle to zero from now on and focus on the real gauge coupling $g$.}. Perturbative superconformal
invariance (the order-by-order vanishing of the gauge beta function and anomalous dimensions for the three scalar fields)
imposes a relation $f(g,\kappa,q,h,N)=0$ between these
parameters, which is generically unknown beyond low loop orders in $g$ 
and where the rank of the gauge group
(taken here to be $\SU(N)$) enters as well. At one loop, the conformal constraint is \cite{Jack:1996qq,Aharony:2002tp}
\be \label{conformalconstraint}
2g^2=\kappa\bar{\kappa}\left[\frac{2}{N^2}(1+q)(1+\qb)+\left(1-\frac{4}{N^2}\right)(1+q\qb+h\hb)\right]\;.
\ee
Taking $q=1,h=0$, the constraint becomes $\kappa=g$ and we recover the $\Ncal=4$ SYM superpotential:
\be \label{N4W}
\Wcal_{\Ncal=4}=g\, {\mathrm{Tr}} \left( \Phi^1 [\Phi^2,\Phi^3]\right) \;.
\ee
The internal symmetry group of $\Ncal=4$ SYM is $\SU(4)_R$, however the $\Ncal=1$ superspace formalism only keeps explicit
a $\SU(3)\times \Urm(1)_R$ subgroup, the $\SU(3)$ being a rotation of the three chiral superfields $\Phi^i$.
The Leigh-Strassler deformation preserves the $\Urm(1)_R$ symmetry, but generically
breaks the $\SU(3)$ to $\Zset_3$ subgroups, which combine to form a non-abelian group known as $\Delta_{27}$ \cite{Aharony:2002hx}. 

Another important special case of the Leigh-Strassler theories arises when $h=0$, but $q\neq 1$. As the
parameter $q$ is often expressed as $q=e^{i\beta}$, this theory has come to be known as the $\beta$-deformation of $\Ncal=4$ SYM.
The $\beta$-deformed theory has additional $\Urm(1)$ symmetries compared to the general case (\ref{LSW}).
This allowed the authors of \cite{Lunin:2005jy} to construct its supergravity dual through a combination of T-duality, shift
in a $\Urm(1)$ angle,
and T-duality back, a procedure known as a TsT transformation. As the angle through which the $\beta$ parameter is introduced is a real
number, this procedure applies only to the real-$\beta$ deformation. However, in \cite{Lunin:2005jy} the dual
background for complex $\beta$ was also obtained, by applying the $\SL(2,R)$ symmetry of IIB supergravity.

The real-$\beta$ case has been extensively studied in the context of planar integrability, starting with \cite{Beisert:2005if},
and its integrability properties turn out to be very similar to $\Ncal=4$ SYM.
For more details and a guide to the literature,
we refer the reader to \cite{Zoubos:2010kh,vanTongeren:2013gva}. On the other hand, the one-loop spin-chain Hamiltonian corresponding to the
complex-$\beta$ deformation does not appear to be integrable \cite{Berenstein:2004ys}. In agreement with this expectation,
string motion on the dual complex-$\beta$ background of \cite{Lunin:2005jy} was shown to be non-integrable \cite{Giataganas:2013dha}.
Furthermore, the one-loop $R$-matrix for the general $(q,h)$
theory was derived and studied in \cite{Bundzik:2005zg} and found \emph{not} to satisfy the Yang-Baxter equation (YBE), apart from special
cases which can mostly be related to the real-$\beta$ deformation through unitary transformations. Since this $R$-matrix will be
the central object of our study, we reproduce it here, in the conventions of \cite{Mansson:2008xv}:
\small  
\be \label{qhRmatrix}
R=\!\frac{1}{1\splus q\qb\splus h\hb}
\left(\begin{array}{ccccccccc}
  \!\!1\splus q\bar{q} \smin h\bar{h}&\hspace{-.3cm}0&\hspace{-.3cm}0&\hspace{-.3cm}0&\hspace{-.3cm}0&\hspace{-.3cm}\!-\!2\bar{h}&\hspace{-.3cm}0&\hspace{-.3cm}2\bar{h}q&\hspace{-.3cm}0\cr 
0&\hspace{-.3cm}2\bar{q}&\hspace{-.3cm}0&\hspace{-.3cm}1\smin q\bar{q}\splus h\bar{h}&\hspace{-.3cm}0&\hspace{-.3cm}0&\hspace{-.3cm}0&\hspace{-.3cm}0&\hspace{-.3cm}2 h\bar{q}\cr 
0&\hspace{-.3cm}0&\hspace{-.3cm}2q&\hspace{-.3cm}0&\hspace{-.3cm}\!-\!2h&\hspace{-.3cm}0&\hspace{-.3cm}q\bar{q}\splus h\bar{h}\smin 1&\hspace{-.3cm}0&\hspace{-.3cm}0\cr 
0&\hspace{-.3cm} q\bar{q}\splus h\bar{h}\smin 1&\hspace{-.3cm}0&\hspace{-.3cm}2q&\hspace{-.3cm}0&\hspace{-.3cm}0&\hspace{-.3cm}0&\hspace{-.3cm}0&\hspace{-.3cm}\!-\!2h\cr 
0&\hspace{-.3cm}0&\hspace{-.3cm}2\bar{h}q&\hspace{-.3cm}0&\hspace{-.3cm}1\splus q\bar{q}\smin h\bar{h}&\hspace{-.3cm}0&\hspace{-.3cm}\!-\!2\bar{h}&\hspace{-.3cm}0&\hspace{-.3cm}0\cr 
2h\bar{q}&\hspace{-.3cm}0&\hspace{-.3cm}0&\hspace{-.3cm}0&\hspace{-.3cm}0&\hspace{-.3cm}2 \bar{q}&\hspace{-.3cm}0&\hspace{-.3cm}1\smin q\bar{q}\splus h\bar{h}&\hspace{-.3cm}0\cr 
0&\hspace{-.3cm}0&\hspace{-.3cm}1\smin q \bar{q}\!+\!h\bar{h}&\hspace{-.3cm}0&\hspace{-.3cm}2h\bar{q}&\hspace{-.3cm}0&\hspace{-.3cm}2\bar{q}&\hspace{-.3cm}0&\hspace{-.3cm}0\cr 
\!-\!2 h&\hspace{-.3cm}0&\hspace{-.3cm}0&\hspace{-.3cm}0&\hspace{-.3cm}0&\hspace{-.3cm}q\bar{q}\splus h\bar{h}\smin 1&\hspace{-.3cm}0&\hspace{-.3cm}2q&\hspace{-.3cm}0\cr
 0&\hspace{-.3cm}\!-\!2\bar{h}&\hspace{-.3cm}0&\hspace{-.3cm}2\bar{h}q&\hspace{-.3cm}0&\hspace{-.3cm}0&\hspace{-.3cm}0&\hspace{-.3cm}0&\hspace{-.3cm}1\splus q\bar{q}\smin h\bar{h}\!\!\cr 
\end{array}\right)\!.
\ee
\normalsize
This $R$-matrix acts on two copies of the vector
space spanned by the three chiral superfields $\Phi^i$, and
is written in the basis $\{11,12,13,21,22,23,31,32,33\}$. We emphasise that this is the $R$-matrix corresponding to
just the holomorphic $\SU(3)$ sector, and not the full ($\SU(4)\simeq \SO(6)$) scalar sector which would also include the
antichiral $\overline{\Phi}_i$ fields.

For $\Ncal=4$ SYM, which corresponds to $q=1,h=0$, the $R$-matrix reduces to just $I_3\otimes I_3$ and reproduces the
XXX Hamiltonian of the schematic form $H_{i,i+1}=I_{i}\otimes I_{i+1}\!-\!P_{i,i+1}$, where $P$ is the permutation matrix. 
For real-$\beta$, the $R$-matrix derived in \cite{Beisert:2005if} (as a special case of the more general $\gamma_i$
deformations \cite{Frolov:2005dj} considered in that work) reduces to the one above in the ``quantum group'' limit of infinite spectral parameter.

Early on, it became clear that the Leigh-Strassler marginal deformations can be thought of as
noncommutative deformations.
The work \cite{Berenstein:2000ux} discussed the noncommutative structure of the moduli space, with the focus mainly being
on non-generic points when $q$ is a root of unity. In the context of twistor strings, it was shown in \cite{Kulaxizi:2004pa}
that a suitable non-anticommutative star product correctly reproduces the amplitudes of the $(q,h)$-deformed
theories at first order in the deformation. However, for generic $(q,h)$ that star product was problematic at higher
orders because of issues with non-associativity. For the real-$\beta$ case, non-associativity is not an issue and
the star product of \cite{Kulaxizi:2004pa} was extended to all orders in \cite{Gao:2006mw}.

In \cite{Lunin:2005jy}, Lunin and Maldacena expressed the real-$\beta$ superpotential in terms of a star product
depending on the $\Urm(1)$ charges of the chiral superfields. The relation to non-commutativity was an important
element in the aforementioned construction of the dual background. The Lunin-Maldacena star product has found many
applications, for instance in relating (planar) $\Ncal=4$ SYM amplitudes to those in the deformed theory \cite{Khoze:2005nd} and
twisting the $\Ncal=4$ SYM Bethe ansatz to its $\beta$-deformed version \cite{Beisert:2005if}.

Similarly to the twistor-string star product, the extension of the Lunin-Maldacena star product to cases beyond the
real-$\beta$ deformation has proven difficult, again because of issues with non-associativity. For the $(q,h)$ cases which
happen to be integrable (in the sense of the $R$-matrix satisfying the YBE) by virtue of their relation to the $\beta$-deformation
through unitary transformations, the gauge-theory
star product was constructed in \cite{Bundzik2007}. A particular such case, termed the $w$-deformation, was further studied in
\cite{Dlamini:2016aaa}, which constructed the star product in terms of a Drinfeld twist (to be reviewed below) and
made progress in understanding how the non-commutativity underlying the gauge theory can appear on the gravity side of the
AdS/CFT correspondence. 

Turning to the general $(q,h)$ theories, progress in understanding their noncommutative structure was made in \cite{Mansson:2008xv},
where, using the FRT relations \cite{FRT90}, it was shown that the superpotential (\ref{LSW}) is invariant under a global
Hopf-algebraic symmetry. In this way, the Leigh-Strassler deformation was understood as a deformation of the
Lie-algebraic $\SU(3)\times \Urm(1)$
R-symmetry group of $\Ncal=4$ SYM to a Hopf algebra. The FRT relations define this algebra at quadratic level, however in
\cite{Mansson:2008xv} associativity was imposed as an additional cubic condition on the generators of this algebra.
Even though this constraint was shown to be consistent at cubic level, the fact that the general $(q,h)$ $R$-matrix does
not satisfy the YBE left open the possibility that the quartic and higher relations will end up trivialising
the algebra. 

In this work, we construct a Drinfeld twist which relates the $(q,h)$ $R$-matrix (\ref{qhRmatrix})
to the (trivial) $R$-matrix of the $\Ncal=4$ SYM
theory. Unlike in \cite{Mansson:2008xv}, we do not force associativity on the algebra, but rather show,
by explicitly constructing the co-associator
of the theory and establishing that the $(q,h)$ R-matrix satisfies the quasi-Hopf Yang-Baxter equation, that the appropriate
algebraic structure is that of a \emph{quasi-Hopf} algebra \cite{Drinfeld90}, which will be reviewed in the next section.

Using our quasi-Hopf Drinfeld twist, we are able to define a star product between the scalar fields of $\Ncal=4$ SYM. As expected,
this star product is non-associative. However, and perhaps surprisingly, the cyclic combinations of fields appearing
in the superpotential of $\Ncal=4$ SYM do not notice the non-associativity. It thus turns out that one can directly obtain
the Leigh-Strassler superpotential (\ref{LSW}) by star-deforming the $\Ncal=4$ SYM superpotential without worrying about associativity.
Expressing the $(q,h)$ superpotential as a star-deformed version of the $\Ncal=4$ SYM superpotential allows us to show that the Leigh-Strassler theories enjoy a global quasi-Hopf symmetry which deforms the $\SU(3)$ part of the $\Ncal=4$ SYM R-symmetry group
in a precise way. So, as long as one is willing to consider quantum group symmetries rather than just Lie algebraic ones,
one finds that the Leigh-Strassler theories have a much higher degree of symmetry than just the naive $\Delta_{27}$.
The prospect of using this hidden symmetry to constrain the observables, and perhaps gain some additional insight on
the supergravity duals of the Leigh-Strassler deformations, certainly provides strong motivation to include quasi-Hopf
algebras in our toolkit in future studies of these theories.

This paper is structured as follows: The next section provides an informal review of aspects of Hopf and quasi-Hopf algebras, with a focus
on Drinfeld twists as a way to generate a new algebra from an existing one. Section 3 shows how the $R$-matrix (\ref{qhRmatrix}) can be
obtained by a Drinfeld twist, while section 4 focuses on the construction of the coassociator, which relates different placements of
parentheses in a quasi-Hopf algebra. In section 5 we use the twist to define suitable star products and show how the superpotential
(\ref{LSW}) can be thought of as a twisted version of (\ref{N4W}). We conclude with a summary of our results and a discussion of
open questions. Several appendices provide further details of our construction.

\section{Drinfeld twists and quasi-Hopf algebras}

In this section we will review, in a very informal way, some general features of Hopf and quasi-Hopf algebras.
For more details and proofs, we refer to textbooks on the subject such as \cite{Majid, ChariPressley}.

\subsection{Hopf algebras}

Recall that the defining feature of an algebra $\Acal$ (over a field $k$, which for us will be $\Cset$)
is an associative product $\cdot$, which takes two copies of the
algebra into a single copy: $\cdot: \Acal\otimes \Acal\ra \Acal$. Similarly, in a coalgebra one has a coproduct
$\Delta$ which takes a single copy to two: $\Delta: \Acal\ra \Acal\otimes \Acal$. In $\Acal$ one also defines
the unit map  $\eta:k\ra \Acal$, which takes $\lambda \ra \lambda 1$, where $\lambda\in k$ and $1$ is the unit element of
$\Acal$ (satisfying $1\cdot X=X\cdot 1=X$ for $X\in \Acal$). In terms of maps, the statement that multiplying an
element by the unit element gives back the original element is expressed as
$\cdot (\eta\otimes \id)=\cdot (\id \otimes \eta)=\id$. Similarly, for a coalgebra one defines
a co-unit $\epsilon:\Acal\ra k$, which expresses the dual of the above statement, i.e. that taking the coproduct of
an element and acting on it by $\epsilon\otimes \id$ or $\id\otimes \epsilon$ should give back the original element:
$(\epsilon\otimes \id)\Delta=(\id\otimes\epsilon)\Delta=\id$. Note that the coproduct in a coalgebra also
needs to be co-associative, which, for $X\in\Acal$, is the statement that
\be
(\id\otimes \Delta)\circ\Delta(X)=(\Delta\otimes \id)\circ\Delta(X)\;.
\ee
A bialgebra arises when combining an algebra and coalgebra such that the product, coproduct, unit and co-unit are all
compatible, which implies that for $X,Y\in\Acal$ one requires the following relations: $\Delta(X\cdot Y)=\Delta(X)\cdot\Delta(Y)$,
$\Delta(1)=1\otimes 1$, $\epsilon(X\cdot Y)=\epsilon(X)\cdot\epsilon(Y)$ and $\epsilon(1)=1$. 

A Hopf algebra is a bialgebra with an additional map $S: \Acal\ra \Acal$ known as the antipode, which is linked to
the existence of an inverse. Unlike $\Delta$ and $\epsilon$, the antipode satisfies $S(X\cdot Y)=S(Y)\cdot S(X)$.

An important case of a Hopf algebra structure arises for the universal enveloping algebra (UEA) of a Lie algebra. This
is the associative algebra obtained by including all polynomials in the Lie algebra elements and imposing
the relations implied by the commutator identities. Focusing on matrix Lie algebras, the product between elements $X$
and $Y$ in the corresponding UEA is just matrix multiplication, and the unit just the unit matrix. The operations
that complete the UEA to a Hopf algebra are
\be \label{Liecoproduct}
\Delta(X)=X\otimes 1+1\otimes X \;,\qquad \;\; \epsilon(X)=0 \;,\qquad \;\; S(X)=-X\;,
\ee
as well as
\be
\Delta(1)=1\otimes 1 \;\qquad \;\; \epsilon(1)=1\;,\qquad\;\;S(1)=1\;.
\ee
One can check that the action of the antipode on a group element corresponds to the inverse group element:
\be
S(e^{\alpha X})=e^{\alpha S(X)}=e^{-\alpha X}\;.
\ee
Note that while the algebra product is of course non-commutative (assuming the underlying Lie algebra is non-abelian),
the above coproduct is co-commutative: Calling $\tau$ the operation of exchanging the two copies of the
algebra, we have $\tau\circ \Delta(X)=\Delta(X)$. This matches our intuition (e.g. from quantum mechanics) of how
Lie algebra elements act on multiparticle states. But the co-commutativity makes the above construction trivial from
the Hopf algebra perspective. Inversely, a non-trivial Hopf algebra structure is one where
both the product is non-commutative and the coproduct non-co-commutative. Of course, any change to the
coproduct has to be done in a way that satisfies the compatibility relations, and, as we will see, a
powerful way of achieving this is through Drinfeld twists. 

Since the $\SU(3)$ part of the R-symmetry group of $\Ncal=4$ SYM is a Lie group, we can consider its
Lie algebra within the above framework, as a trivial Hopf algebra with the product being the usual matrix
product and the coproduct being the co-commutative one given in (\ref{Liecoproduct}). This is the Hopf
algebra that we will be Drinfeld-twisting later on.

\subsection{Quasitriangular Hopf algebras}

We will further specialise to quasitriangular Hopf algebras, where there exists an
element $R\in \Acal\otimes \Acal$, called the universal $R$-matrix. This matrix governs the
non-cocommutativity of the algebra, in the sense that the opposite coproduct, while not equal
to the coproduct as in the above example, is related to it via conjugation with $R$: 
\be \label{opDelta}
\tau\circ \Delta(X)=R \Delta(X) R^{-1}\;, \;\quad X \in \Acal\;.
\ee
The matrix $R$ is also taken to satisfy the conditions:
\be \label{quasitrig}
(\Delta\otimes \id)(R)=R_{13}R_{23} \;,\quad\; (\id \otimes \Delta)(R)=R_{13}R_{12}\;,
\ee
which, together with (\ref{opDelta}), can be shown to imply the Yang-Baxter equation
\be \label{QYBE}
R_{12}R_{13}R_{23}=R_{23}R_{13}R_{12}\;.
\ee
A special case arises when the $R$-matrix further satisfies the condition
\be \label{triangular}
R_{21}=R_{12}^{-1}\;.
\ee
In this case the Hopf algebra is called \emph{triangular}.

Clearly, the co-commutative UEA of a Lie algebra is a triangular Hopf algebra with trivial $R$-matrix: $R=1\otimes 1$.

The above discussion in terms of the universal $R$-matrix was independent of a choice of representation for the
algebra. However, the $R$-matrix (\ref{qhRmatrix}) that we are interested in is evaluated in the product
of the fundamental representation, so our discussion in the following (and all our results for the twists) will
only apply in this limited setting. (In general, the construction of universal $R$-matrices and corresponding twists
is a difficult problem, even for well-studied quantum groups.)

\subsection{Drinfeld Twists}

We will now consider relations between inequivalent Hopf algebras which arise due to \emph{Drinfeld twists} \cite{Drinfeld90}.
We will remain within the quasitriangular case as above, but eventually restrict to the triangular case which will be relevant
to our case. 

Given an initial quasitriangular Hopf algebra, with $R$-matrix $R_0$, a Drinfeld twist is an element $F \in \Acal\otimes \Acal$, which is
invertible and satisfies $(\epsilon\otimes \id)F=(\id\otimes \epsilon)F=1$, known as the co-unital condition. The twist
can be used to deform the coproduct of the algebra as
\be \label{twistedcoproduct}
\Delta_F(X)=F \Delta(X)F^{-1} \;,
\ee
and its $R$-matrix by
\be
R^F_{12}=F_{21} R_{12} F_{12}^{-1}\;.
\ee \label{twistedR}
Twists preserve quasitriangularity, as well as triangularity. We can easily check the latter:
\be
R^F_{21}=\tau\circ(F_{21}R_{12}F_{12}^{-1})=F_{12}R_{21}F_{21}^{-1}=F_{12}R_{12}^{-1}F_{21}^{-1}=(R^F)^{-1}_{12}\;.
\ee
 In our case, the
 initial $R$-matrix is that of $\Ncal=4$ SYM, which, as discussed in the introduction, is
 just the identity matrix $I\otimes I$.\footnote{We will use $I$ instead of $1$
   for the identity when we are specifically referring to the $3\times3$ identity matrix.} So for $R_{q,h}$ in (\ref{qhRmatrix}), the twist relation (\ref{twistedR}) reduces to
\be \label{factorisingtwist}
R_{q,h}=F_{21}F_{12}^{-1}\;.
\ee
This is known as a factorising twist (see e.g. \cite{Maillet:1996yy}). Clearly an $R$-matrix constructed
as in (\ref{factorisingtwist}) will automatically satisfy the triangular condition (\ref{triangular}), and it is shown in
\cite{Drinfeld90} that any triangular $R$-matrix admits a factorising twist.

If a twist additionally satisfies the 2-cocycle condition
\be \label{cocycle}
(F\otimes 1) (\Delta \otimes \id)(F)=(1\otimes F)(\id \otimes \Delta)(F) \;,
\ee
(with $\Delta$ the undeformed coproduct) then twisting a Hopf algebra produces another Hopf algebra. 
This is the setting of the $w$-deformation considered in \cite{Dlamini:2016aaa}. 
To distinguish the Drinfeld twists satisfying (\ref{cocycle}) from the general case, they
were termed \emph{Hopf twists} in \cite{Dlamini:2016aaa}. When twisting a quasitriangular Hopf algebra, the
cocycle condition guarantees that the $R$-matrix of the twisted theory is also quasitriangular.

Drinfeld twists have found numerous applications in the study of quantum groups, although their
applicability is often limited by the lack of explicit expressions for the twists. In the context of
integrability, Drinfeld twists have been advocated as a powerful tool within the algebraic Bethe ansatz
approach in \cite{Maillet:1996yy}.  In the more specific context of AdS/CFT integrability, the
$\gamma_i$-twisted Bethe equations of \cite{Beisert:2005if} have been derived via an (abelian) Drinfeld
twist in \cite{Ahn:2010ws}. In \cite{vanTongeren:2015uha}, Drinfeld twists were used to achieve a
better understanding of Yang-Baxter deformations of the string sigma model and their field theory interpretation.

In our current setting, since we know that the YBE is not satisfied for the general $(q,h)$ $R$-matrix, we can
infer that the Drinfeld twist factorising $R_{q,h}$ as in (\ref{factorisingtwist})
cannot satisfy the cocycle condition. Relaxing (\ref{cocycle})
brings us into the realm of quasi-Hopf algebras, which we turn to in the next section.

\subsection{Quasitriangular Quasi-Hopf algebras}

The definition of quasi-Hopf algebras involves the same ingredients as Hopf algebras, i.e. an associative algebra,
and a coalgebra given by a coproduct and co-unit, but brings in a new item, an element
$\Phi\in\Acal\otimes \Acal \otimes \Acal$ called the coassociator.\footnote{We hope that no confusion will arise between $\Phi$ and the chiral superfields $\Phi^i$, which always
  carry indices.} Here the coproduct is allowed to be non-co-associative, which brings us out of the domain of Hopf algebras.
However, similarly to how the $R$-matrix in a quasitriangular Hopf algebra controls the non-commutativity
of the coproduct, the coassociator controls its non-associativity:\footnote{We note that the conventions for $\Phi$ vary in
  the literature. Ours are as in \cite{Drinfeld90,Majid} but differ from \cite{ChariPressley} by $\Phi\rightarrow\Phi^{-1}$.}
\be \label{nonassoc}
(\id\otimes \Delta)\circ\Delta(X)=\Phi [(\Delta\otimes \id)\circ \Delta(X)] \Phi^{-1}\;.
\ee
 We see that the two placements of parentheses (``bracketings'') are equivalent by conjugation with $\Phi$.
This is often described as a quasi-associative condition, as it corresponds to breaking co-associativity
in the weakest possible sense: Assuming one knows $\Phi$, one can always use it to express a term with a given
bracketing as a linear combination of terms with the other bracketing and thus
ensure that the placement of parentheses is the same for all terms in an equation. For this reason, $\Phi$ is also
often called a \emph{re-associator}.

The coassociator identity (\ref{nonassoc}) also has implications for the module/representation
space of our algebra, which we call $\Vcal$. We denote the action of an element $X$ of $\Acal$ on an element
$a \in \Vcal$ as $a'=X\trr a$. To act with $X$ on the tensor product $\Vcal\otimes \Vcal$ we need to use the
coproduct, $(a'\otimes b')=\Delta(X)\trr [a\otimes b]$. If we now wish to act on the cubic tensor product $V\otimes (V\otimes V)$,
  we need to iterate the coproduct in a way that matches the bracketing:
$a'\otimes (b'\otimes c') =(\id\otimes \Delta)\Delta(X) \trr [a\otimes(b\otimes c)]$.
Similarly, the action on $(V\otimes V)\otimes V$ will be 
$ (a'\otimes b')\otimes c')=(\Delta\otimes \id)\Delta(X) \trr [(a\otimes b)\otimes c]$. Given
(\ref{nonassoc}), covariance of the action of
$\Acal$ on $\Vcal$ implies that the left- and right- cubic elements need to be related as
$a\otimes(b\otimes c)=\Phi\,\trr (a\otimes b)\otimes c$, since then we can compute 
\be
\begin{split}
a'\otimes (b'\otimes c') &=[(\id\otimes \Delta)\circ\Delta(X)] \trr [a\otimes(b\otimes c)]\\&=
\Phi [(\Delta\otimes \id)\circ\Delta(X)] \Phi^{-1}\, \trr\, \Phi \, \trr \, [(a\otimes b)\otimes c]
=\Phi\,\trr (a'\otimes b')\otimes c'\;.
\end{split}
\ee
We conclude that the left- and right- placements of parentheses for the module tensor product are related by the
coassociator as
\be \label{assocmorphism}
\Vcal\otimes (\Vcal\otimes \Vcal)=\Phi\, (\Vcal\otimes \Vcal) \otimes \Vcal\;.
\ee
This associativity isomorphism can be thought of as mapping between the two nodes of the one-dimensional 
associahedron $K_3$. Recall that an associahedron is the polytope where each node corresponds to an inequivalent
placement of parentheses. Since for three elements we only have two possible placements, the
associahedron $K_3$ has just two nodes which we can label ${\bf L}$ and ${\bf R}$ based on the placement
of the parenthesis. So we can write (\ref{assocmorphism}) as $\Phi: {\bf L}\rightarrow {\bf R}$. 

  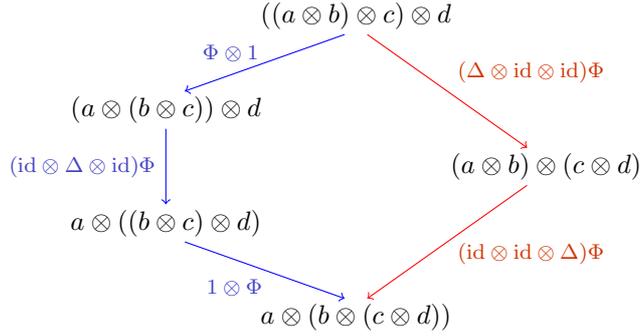
\begin{figure}[t]
\begin{center}
    \begin{tikzpicture}[scale=0.5]
      \node at (5,0) {\small $\left((a\otimes b)\otimes c\right)\otimes d$};
      \node at (0,-2.5) {\small $\left(a\otimes (b\otimes c)\right)\otimes d$};
      \node at (0,-5.5) {\small $a\otimes \left((b\otimes c)\otimes d\right)$};
      \node at (10,-4) {\small $(a\otimes b)\otimes (c\otimes d)$};
      \node at (5,-8) {\small $a\otimes\left( b\otimes (c\otimes d)\right)$};
 \draw[->,blue] (4.7,-.5) --(0.5,-2);      
 \draw[->,blue] (0,-3) --(0,-5);
 \draw[->,blue] (0.5,-6) --(4.7,-7.5);
 \draw[->,red] (5.3,-.5) --(9.5,-3.5);
 \draw[->,red] (9.5,-4.5) --(5.3,-7.5);
 \node at (9.6,-1.5) {\scriptsize\cdr $(\Delta\otimes\id\otimes \id)\Phi$};
 \node at (9.6,-6.3) {\scriptsize\cdr $(\id\otimes \id\otimes\Delta)\Phi$};
 \node at (1.7,-1) {\scriptsize\cdb $\Phi\otimes 1$};
 \node at (-2.2,-4) {\scriptsize\cdb $(\id\otimes \Delta\otimes \id)\Phi$};
   \node at (1.8,-7.2) {\scriptsize\cdb $1\otimes\Phi$};
    \end{tikzpicture}
    \caption{A graphical depiction of the pentagon identity, showing the maps between the different placements of parentheses for
      four elements $a,b,c,d$ of $\Vcal$.}\label{pentagonfig}
      \end{center}
  \end{figure}

For four elements, there are five inequivalent bracketings, and the corresponding associahedron $K_4$ is a pentagon.
As depicted graphically in Fig. \ref{pentagonfig}, there are two possible ways to construct maps between two given
vertices of this pentagon. Since these two routes must agree, one derives the following 
consistency check on the coassociator, known as the pentagon identity:
\be \label{pentagon}
(\id\otimes \id\otimes \Delta)(\Phi)(\Delta\otimes \id\otimes \id)(\Phi)=(1\otimes \Phi)(\id\otimes \Delta\otimes \id)(\Phi)(\Phi\otimes 1)\;.
\ee
After these remarks on the role of the coassociator, let us return to our definition of a quasi-Hopf algebra. The structure
we have described so far is a \emph{quasi-bialgebra}. For a quasi-Hopf algebra we also need to introduce the
analogue of an antipode. It turns out that in the quasi-Hopf case this notion is provided by a triple $(S,\alpha,\beta)$,
sometimes termed a quasi-antipode, containing the antipode together with two canonical elements
$\alpha$ and $\beta$. As we will not require more details of the quasi-antipode
construction in this work, we refer to \cite{Drinfeld90} and the literature for further details.

For our purposes, we will think of quasi-Hopf algebras as the algebraic structure arising when Drinfeld-twisting
Hopf algebras  by twists that do not satisfy the 2-cocycle condition (\ref{cocycle}). (The twists still need
to be invertible and co-unital). In this sense, quasi-Hopf algebras are simpler than
Hopf algebras, as they allow for a larger ``gauge invariance'' given by the Drinfeld twists.

In terms of the twist, the coassociator is expressed in terms of that of the untwisted algebra as:
\be \label{coassociator}
\Phi=F_{23} (\id\otimes \Delta) (F) \Phi_0 (\Delta\otimes \id) (F^{-1}) F^{-1}_{12} \;.
\ee
If the untwisted algebra is a Hopf algebra, with trivial coassociator $\Phi_0=1\otimes 1\otimes 1$, we
can read the 2-cocycle condition as the statement that the coassociator remains trivial after twisting.
This guarantees that the twisted Hopf algebra is still an (associative) Hopf algebra. However, since in our
case we will be interested in more general twists, we will be working within the class of (quasitriangular)
quasi-Hopf algebras, and thus expect (\ref{coassociator}) to lead to a non-trivial coassociator.

\begin{figure}[t]
  \begin{center}
    \begin{tikzpicture}[scale=0.5]
            \node at (1,.7) {\small $1$};             \node at (4,.7) {\small $2$};            \node at (7,.7) {\small $3$};
      \node[style={circle, fill=blue, inner sep=0pt, minimum size=1mm}] at (1,0) () {};
            \node[style={circle, fill=\darkgreen, inner sep=0pt, minimum size=1mm}] at (4,0) () {};
            \node[style={circle, fill=red, inner sep=0pt, minimum size=1mm}] at (7,0) () {};
      \draw[-,blue] (1,0) -- (2.3,-1.3);
            \draw[-,blue] (2.7,-1.7) -- (4,-3) --(4,-3.5);
            \draw[-,\darkgreen] (4,0) --(1,-3) -- (1,-3.5);
            \draw[-,\darkgreen,densely dotted] (1,-3.5) --(1,-7.5);
            \draw[-,blue,densely dotted] (4,-3.5) -- (4,-4) -- (5.3,-5.3);
            \draw[-,red] (7,0) -- (7,-3.5);
            \draw[-,red,densely dotted] (7,-3.5) -- (7,-4) -- (4,-7) -- (4,-7.5);
            \draw[-,blue,densely dotted] (5.7,-5.7) -- (7,-7) --(7,-7.5);
            
            \draw[-,\darkgreen] (1,-7.5) -- (1,-8) -- (2.3,-9.3);
            \draw[-, \darkgreen] (2.7,-9.7) -- (4,-11) -- (4,-11.5);
            \draw[-,red] (4,-7.5) -- (4,-8) -- (1,-11) -- (1,-11.5);
            \draw[-,blue] (7,-7.5) -- (7,-11.5);
            \draw[-,blue,densely dotted] (7,-11.5) -- (7,-12);
            \draw[-,red,densely dotted] (1,-11.5) -- (1,-12);
            \draw[-,\darkgreen,densely dotted] (4,-11.5) -- (4,-12);
            \draw[->,\graphite] (0,-3.5) -- (8,-3.5);
            \draw[->,\graphite] (8,-7.5) -- (0,-7.5);
            \draw[->,\graphite] (0,-11.5) -- (8,-11.5);
            
            \node[style={circle, fill=blue, inner sep=0pt, minimum size=1mm}] at (7,-12) () {};
            \node[style={circle, fill=\darkgreen, inner sep=0pt, minimum size=1mm}] at (4,-12) () {};
            \node[style={circle, fill=red, inner sep=0pt, minimum size=1mm}] at (1,-12) () {};
            \node at (1,-12.7) {\small $3$};             \node at (4,-12.7) {\small $2$};            \node at (7,-12.7) {\small $1$};
            \node at (-1,-1.5) {\scriptsize $R_{12}$};
                \node at (-1,-3.5) {\scriptsize $\Phi_{213}$};
                \node at (-1,-5.5) {\scriptsize $R_{13}$};
                \node at (-1,-7.5) {\scriptsize $\Phi^{-1}_{231}$};
                \node at (-1,-9.5) {\scriptsize $R_{23}$};
                \node at (-1,-11.5) {\scriptsize $\Phi_{321}$};

                \node at (9.5,-6) {$=$};

                \node at (12,.7) {\small $1$};             \node at (15,.7) {\small $2$};            \node at (18,.7) {\small $3$};
      \node[style={circle, fill=blue, inner sep=0pt, minimum size=1mm}] at (12,0) () {};
            \node[style={circle, fill=\darkgreen, inner sep=0pt, minimum size=1mm}] at (15,0) () {};
            \node[style={circle, fill=red, inner sep=0pt, minimum size=1mm}] at (18,0) () {};
            \draw[-,blue] (12,0) -- (12,-0.5);
            \draw[-,\darkgreen] (15,0) -- (15,-0.5);
            \draw[-,red] (18,0) -- (18,-0.5);
            \draw[-,blue,densely dotted] (12,-0.5) -- (12,-4.5);
            \draw[-,\darkgreen,densely dotted] (15,-0.5) -- (15,-1) -- (16.3,-2.3);
            \draw[-,\darkgreen,densely dotted] (16.7,-2.7) -- (18,-4) -- (18,-4.5);
            \draw[-,red,densely dotted] (18,-0.5) -- (18,-1) -- (15,-4) -- (15, -4.5);
            \draw[-,blue] (12,-4.5) -- (12,-5) -- (13.3,-6.3);
            \draw[-,blue] (13.7,-6.7) -- (15,-8) -- (15,-8.5);
            \draw[-,red] (15,-4.5) -- (15,-5) -- (12,-8) -- (12,-8.5);
            \draw[-,\darkgreen] (18,-4.5) -- (18,-8.5);
            \draw[-,red,densely dotted] (12,-8.5) -- (12,-12);
            \draw[-,blue,densely dotted] (15,-8.5) -- (15,-9) -- (16.3,-10.3);
            \draw[-,blue,densely dotted] (16.7,-10.7) -- (18,-12);
            \draw[-,\darkgreen,densely dotted] (18,-8.5) -- (18,-9) -- (15,-12);
            \draw[->,\graphite] (11,-0.5) -- (19,-0.5);
            \draw[->,\graphite] (19,-4.5) -- (11,-4.5);
            \draw[->,\graphite] (11,-8.5) -- (19,-8.5);            
            \node[style={circle, fill=blue, inner sep=0pt, minimum size=1mm}] at (18,-12) () {};
            \node[style={circle, fill=\darkgreen, inner sep=0pt, minimum size=1mm}] at (15,-12) () {};
            \node[style={circle, fill=red, inner sep=0pt, minimum size=1mm}] at (12,-12) () {};
            \node at (12,-12.7) {\small $3$};             \node at (15,-12.7) {\small $2$};            \node at (18,-12.7) {\small $1$};
 \node at (20,-0.5) {\scriptsize $\Phi_{123}$};
            \node at (20,-2.5) {\scriptsize $R_{23}$};
                \node at (20,-4.5) {\scriptsize $\Phi^{-1}_{132}$};
                \node at (20,-6.5) {\scriptsize $R_{13}$};
                \node at (20,-8.5) {\scriptsize $\Phi_{312}$};
                \node at (20,-10.5) {\scriptsize $R_{12}$};                           
    \end{tikzpicture}
  \caption{A graphical representation of the quasi-Hopf Yang-Baxter equation. In order for the $R$-matrix to be able to interchange two particles, they need to live on the appropriate node of the 1-d associahedron. Solid lines indicate the node ${\bf L}: (\Vcal\otimes \Vcal)\otimes \Vcal$, while
    dashed ones the node ${\bf R}: \Vcal\otimes (\Vcal\otimes \Vcal)$. An arrow pointing to the right indicates the map $\Phi: {\bf L}\ra {\bf R}$,
    while an arrow to the left indicates $\Phi^{-1}: {\bf R}\ra {\bf L}$. Read from top to bottom, both the left and right diagrams map $(\Vcal_1\otimes \Vcal_2)\otimes \Vcal_3$ to
  $\Vcal_3\otimes (\Vcal_2\otimes \Vcal_1)$, and the qHYBE is the statement that these two paths are equal.} \label{qHYBEfig}
    \end{center}
  \end{figure}
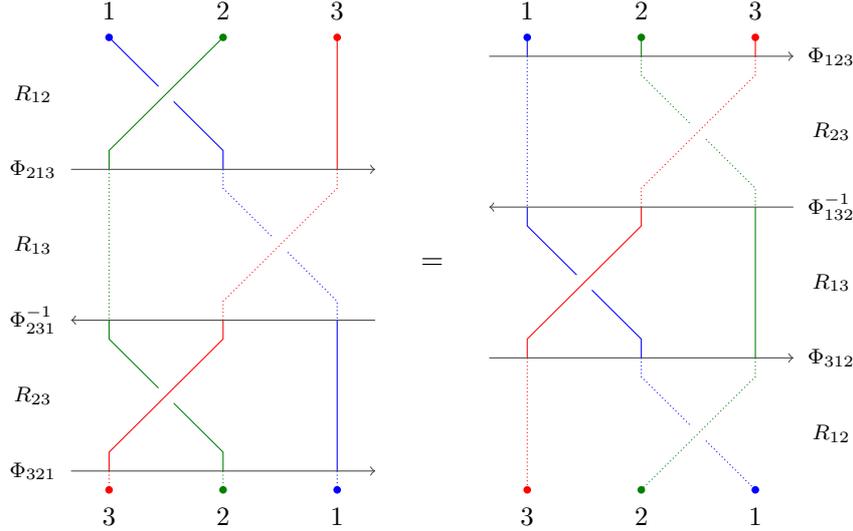

The additional structure coming with quasitriangularity also extends naturally to quasi-Hopf algebras.
In a quasitriangular quasi-Hopf algebra, the role of the $R$-matrix in relating the coproduct and opposite coproduct
remains the same (\ref{opDelta}), however the relations (\ref{quasitrig}) are modified to
\be
(\Delta\otimes \id)(R)=\Phi_{312}R_{13}\Phi_{132}^{-1}R_{23}\Phi_{123} \;\;\text{and}\;\;
(\id\otimes \Delta)(R)=\Phi_{231}^{-1} R_{13} \Phi_{213} R_{12} \Phi_{123}^{-1}\;.
\ee
These relations imply that, instead of the usual YBE (\ref{QYBE}) the $R$-matrix satisfies the
quasi-Hopf Yang-Baxter equation (qHYBE):\footnote{This condition is often denoted just ``quasi-YBE'' in the literature, and
  also holds in more general situations, such as the weak quasitriangular quasi-Hopf case of \cite{Mack:1991tg}. We will use
  qHYBE to emphasise our quasi-Hopf setting.}
\be \label{qhYBE}
R_{12}\Phi_{312}R_{13}\Phi^{-1}_{132} R_{23}\Phi_{123}=\Phi_{321}R_{23}\Phi^{-1}_{231}R_{13}\Phi_{213}R_{12}\;.
\ee 
One can depict this relation graphically as in Fig. \ref{qHYBEfig}, where as usual one thinks of the $R$-matrix as a
braiding element, but where one also has to account for the need to map between the two nodes of the
associahedron in order for its action on the appropriate vector spaces (which we think of physically as
the states corresponding to three particles which are scattering) to make sense.\footnote{If one were to consider
  quasi-Hopf scattering of four particles, one would need to take into account the associativity maps between the five
  nodes of the pentagon (Fig. \ref{pentagonfig}), and so on for more particles.}

 The question of whether quasi-Hopf symmetries can appear in Quantum Field Theory
 was addressed in \cite{Mack:1991tg}, where it was shown that (quasitriangular) quasi-Hopf algebras (as well as a generalisation
 called weak quasi-Hopf) are compatible with all our usual expectations from QFT. So far, most physical applications of Quasi-Hopf
 algebras have been
in the context of 2-d CFT (e.g. in relation to the Knizhnik-Zamolodchikov equation, as reviewed in \cite{ChariPressley},
see also \cite{Gomez96}). Interestingly, it also turns out \cite{Babelon:1995rz} that the dynamical Yang-Baxter equation
can be obtained by a quasi-Hopf Drinfeld twist of a particular type, see \cite{Jimbo:1999zz} for a discussion. 
Recently, quasi-Hopf algebras have also made an appearance in target-space string theory \cite{Mylonas:2013jha,
  Barnes:2016cjm}, in the context of non-geometric compactifications, where the geometry perceived by closed strings
can be non-associative (see \cite{Plauschinn:2018wbo} for a recent review). The pentagon identity that we reviewed
also arises in this context, see \cite{Bakas:2013jwa} for a discussion.

In the following sections we will show how to twist the $\Ncal=4$ SYM $R$-matrix $R=I\otimes I$ in order to obtain
the $R$-matrices relevant to the Leigh-Strassler deformations, and conclude that the resulting structure fits precisely
into the framework above.

\section{The Leigh-Strassler $R$-matrix from a twist}

Let us now focus on the $R$-matrix (\ref{qhRmatrix}). As mentioned, this $R$-matrix
does not generically satisfy the Quantum Yang-Baxter equation (\ref{QYBE}). In the context of the FRT relations
\cite{FRT90}, which were studied in this context in \cite{Mansson:2008xv}, failure of the YBE can lead to the trivialisation of
the algebra at higher orders if associativity is enforced. In this work we will argue that a more fruitful approach might
be to not enforce associativity and instead work in the wider context of quasi-Hopf algebras.

Among several interesting properties of the $(q,h)$- $R$-matrix, it was observed in \cite{Dlamini:2016aaa} that
it is \emph{triangular}, satisfying the relation (\ref{triangular}), or in indices:
\be \label{triangularindices}
R^{j\;i}_{\;l\;k}=(R^{-1})^{i\;j}_{\;k\;l}\;\quad \Rightarrow \quad \;
R^{n\;m}_{ \;j\;i} R^{i\;j}_{\;k\;l}=\delta^m_{\;k}\delta^n_{\;l}\;.
\ee
The triangular property suggests that $R_{q,h}$ arises from a Drinfeld twist. Such twists were constructed
in \cite{Dlamini:2016aaa} for special cases where $R_{q,h}$ satisfies the YBE, in particular the
real-$\beta$ deformation (where $\qb=1/q,h=\hb=0$) and the $w$-deformation (where $q=\qb=1+w,h=\hb=w$). Those twists
have several special properties, and in particular they are abelian, in the sense that they can be expressed as
exponentials of commuting matrices (Cartan matrices for the real-$\beta$ case and shift matrices for the $w$-deformation). 
In the next section we will extend the twists of \cite{Dlamini:2016aaa} to the general (non-abelian) case.

The $(q,h)$ $R$-matrix is also unitary as a $9\times 9$ matrix, which together with the triangular property implies that it
is of \emph{real type}, i.e.
\be \label{Rrealtype}
\overline{R^{i\;j}_{\;k\;l}}=R^{l\;k}_{\;j\;i}\;.
\ee
This will play a role in section \ref{MixedRelations}, when we discuss the mixed quantum plane relations arising from the $R$-matrix. 

Finally, one can check that the $R$-matrix (\ref{qhRmatrix}) has a $\Zset_3$ invariance: for any choice of indices,
$R^{i\;j}_{\;k\;l}=R^{i+1\;j+1}_{\;k+1\;l+1}=R^{i-1\;j-1}_{\;k-1\;l-1}$. This of course reflects the $\Zset_3$ symmetry of the superpotential,
$\Phi^1\ra\Phi^2\ra\Phi^3\ra \Phi^1$, which is part of the $\Delta_{27}$ discrete symmetry group of the theory. All our expressions
in the following will need to respect this $\Zset_3$ invariance.

\subsection{The imaginary $\beta$ twist}

As a warm-up, and to motivate the construction of the $(q,h)$-twist,
let us consider the case of imaginary $\beta$, i.e. $q=e^{i\beta}$ real. We would like to construct
the twist which leads to the imaginary-$\beta$ R-matrix, in other words it should satisfy (\ref{factorisingtwist})
for $\qb=q$ and $h=\hb=0$. 

Our approach will be to suitably exponentiate the classical twist. For imaginary $\beta$ the classical $r$-matrix is\footnote{Although
  in this case the $R$-matrix is real, in general it is unitary so our 
  convention is to expand it as $R_{12}=I\otimes I+i r_{12}+\cdots$ so that the classical $r$-matrix is hermitian. Similarly, we will
  write $F_{12}=I\otimes I+i f_{12}+\cdots$. }
\be \label{rib}
r=-(q-1)\begin{pmatrix}0&0&0&0&0&0&0&0&0\cr 0&0&0&-i&0&0&0&0&0\cr 
 0&0&0&0&0&0&i&0&0\cr 0&i&0&0&0&0&0&0&0\cr 0&0&0&0&0&0&0&0&0\cr 0
 &0&0&0&0&0&0&-i&0\cr 0&0&-i&0&0&0&0&0&
 0\cr 0&0&0&0&0&i&0&0&0\cr 0&0&0&0&0&0&0&0&0\cr 
\end{pmatrix}\;.
\ee
Let us assume that, like the $R$-matrix, the twist is triangular, $F^{-1}_{12}=F_{21}$. Then the classical twist,
which satisfies (\ref{factorisingtwist}) to first order in $(q-1)$, is just
$f_{12}=-\half r_{12}$. It is easy to see that simply exponentiating $f_{12}$ does not lead to
a twist satisfying (\ref{factorisingtwist}). Fortunately, it turns out that the next best guess
works: Let us start by writing $f_{12}=\frac{q-1}{2} \tilde{f}_{12}$, where 
$\tilde{f}_{12}$ is the matrix in (\ref{rib}) without the prefactor. We then replace the $(q-1)/2$ coefficient by a to-be-determined
function $\alpha(q)$, which should reduce to $(q-1)/2$ at first order. Exponentiating the matrix
$\alpha(q)\tilde{f}_{12}$, we find:
\begin{equation}
 F_{12}=e^{i\alpha {\tilde f}_{12}}=\begin{pmatrix}1&0&0&0&0&0&0&0&0\cr 0&\cos \alpha&0&\sin \alpha&0&0&0&0&0\cr 0&0&
 \cos \alpha&0&0&0&-\sin \alpha&0&0\cr 0&-\sin \alpha&0&\cos \alpha&0&0&0&0&0\cr 0&0&0&0&
 1&0&0&0&0\cr 0&0&0&0&0&\cos \alpha&0&\sin \alpha&0\cr 0&0&\sin \alpha&0&0&0&\cos \alpha&
 0&0\cr 0&0&0&0&0&-\sin \alpha&0&\cos \alpha&0\cr 0&0&0&0&0&0&0&0&1\cr 
\end{pmatrix}.
\end{equation}
Requiring (\ref{factorisingtwist}) as well as the appropriate limits, we 
easily obtain $\cos \alpha=\frac{1+q}{\sqrt{2(q^2+1)}}$. Inverting
these relations we find
\be \label{alphaq}
\alpha(q)=\arccos\left(\frac{1+q}{\sqrt{2(q^2+1)}}\right)=\frac{q-1}{2}-\frac{(q-1)^2}{4}+\cdots
\ee

which agrees with our requirement that $F_{12}$ reduces to the classical twist at first order.
In conclusion, the imaginary-$\beta$ twist we obtain is:
\be \label{imbF}
F=\begin{pmatrix}1&0&0&0&0&0&0&0&0\cr 0&{{q+1}\over{\sqrt{2}\,\sqrt{q^2+1}
 }}&0&{{q-1}\over{\sqrt{2}\,\sqrt{q^2+1}}}&0&0&0&0&0\cr 0&0&{{q+1
 }\over{\sqrt{2}\,\sqrt{q^2+1}}}&0&0&0&-{{q-1}\over{\sqrt{2}\,\sqrt{q
 ^2+1}}}&0&0\cr 0&-{{q-1}\over{\sqrt{2}\,\sqrt{q^2+1}}}&0&{{q+1
 }\over{\sqrt{2}\,\sqrt{q^2+1}}}&0&0&0&0&0\cr 0&0&0&0&1&0&0&0&0\cr 0&
 0&0&0&0&{{q+1}\over{\sqrt{2}\,\sqrt{q^2+1}}}&0&{{q-1}\over{\sqrt{2}
 \,\sqrt{q^2+1}}}&0\cr 0&0&{{q-1}\over{\sqrt{2}\,\sqrt{q^2+1}}}&0&0&0
 &{{q+1}\over{\sqrt{2}\,\sqrt{q^2+1}}}&0&0\cr 0&0&0&0&0&-{{q-1}\over{
 \sqrt{2}\,\sqrt{q^2+1}}}&0&{{q+1}\over{\sqrt{2}\,\sqrt{q^2+1}}}&0
 \cr 0&0&0&0&0&0&0&0&1\cr 
\end{pmatrix}.
\ee
This twist has unit determinant by construction, and reduces to
the identity as $q\rightarrow 1$.

\subsection{The general $(q,h)$--twist}

We now turn to the twist leading to the general $R$-matrix (\ref{qhRmatrix}). Now the
classical twist will depend on the four parameters $(q-1),(\qb-1),h,\hb$. Instead of exponentiating
the sum of these four matrices as above, we take a shortcut to the form of the full twist. 
Guided by the form of the $R$-matrix, we parametrise the twist as follows:
\be \label{qhtwist}
F_{12}=\begin{pmatrix}a&0&0&0&0&e&0&f&0\cr 0&b&0&c&0&0&0&0&g\cr 0&0&i&0&j&0&d&0&
 0\cr 0&d&0&i&0&0&0&0&j\cr 0&0&f&0&a&0&e&0&0\cr g&0&0&0&0&b&0&c&0\cr 
 0&0&c&0&g&0&b&0&0\cr j&0&0&0&0&d&0&i&0\cr 0&e&0&f&0&0&0&0&a\cr \end{pmatrix}\;,
\ee
where the non-zero coefficients are compatible with the $\Zset_3$ symmetry of the
theory. Requiring (\ref{factorisingtwist}), unit determinant, invertibility and the appropriate limits, and
after some computation, we find that the coefficients are given by the following expressions:

\small
\be \label{aparam}
a={{\sqrt{2}\,h\,{\it \bar{h}}\,\sqrt{q+1}\,\sqrt{{\it \bar{q}}+1}}\over{\sqrt{
 q\,{\it \bar{q}}+h\,{\it \bar{h}}+1}\,\left(q\,{\it \bar{q}}-{\it \bar{q}}-q+2\,h\,
 {\it \bar{h}}+1\right)}}+{{\left(q-1\right)\,\left({\it \bar{q}}-1\right)
 }\over{q\,{\it \bar{q}}-{\it \bar{q}}-q+2\,h\,{\it \bar{h}}+1}}\;,
\ee

\be
b={{h\,{\it \bar{h}}}\over{q\,{\it \bar{q}}-{\it \bar{q}}-q+2\,h\,{\it \bar{h}}+1}}+{{q^2
 \,{\it \bar{q}}^2-q\,{\it \bar{q}}^2+q^2\,{\it \bar{q}}+h\,{\it \bar{h}}\,q\,{\it \bar{q}}-h
 \,{\it \bar{h}}\,{\it \bar{q}}-{\it \bar{q}}-2\,q^2+3\,h\,{\it \bar{h}}\,q+q+h\,{\it \bar{h}}
 +1}\over{\sqrt{2}\,\sqrt{q+1}\,\sqrt{{\it \bar{q}}+1}\,\sqrt{q\,{\it \bar{q}}+
 h\,{\it \bar{h}}+1}\,\left(q\,{\it \bar{q}}-{\it \bar{q}}-q+2\,h\,{\it \bar{h}}+1\right)
 }}\;,
\ee

\be
c={{h\,{\it \bar{h}}}\over{q\,{\it \bar{q}}-{\it \bar{q}}-q+2\,h\,{\it \bar{h}}+1}}+{{q^2
 \,{\it \bar{q}}^2-q\,{\it \bar{q}}^2-q^2\,{\it \bar{q}}+h\,{\it \bar{h}}\,q\,{\it \bar{q}}-h
 \,{\it \bar{h}}\,{\it \bar{q}}+{\it \bar{q}}-h\,{\it \bar{h}}\,q+q-3\,h\,{\it \bar{h}}-1
 }\over{\sqrt{2}\,\sqrt{q+1}\,\sqrt{{\it \bar{q}}+1}\,\sqrt{q\,{\it \bar{q}}+h
 \,{\it \bar{h}}+1}\,\left(q\,{\it \bar{q}}-{\it \bar{q}}-q+2\,h\,{\it \bar{h}}+1\right)
 }}\;,
\ee

\be
d={{h\,{\it \bar{h}}}\over{q\,{\it \bar{q}}-{\it \bar{q}}-q+2\,h\,{\it \bar{h}}+1}}-{{q^2
 \,{\it \bar{q}}^2-q\,{\it \bar{q}}^2-q^2\,{\it \bar{q}}+3\,h\,{\it \bar{h}}\,q\,{\it \bar{q}}
 +h\,{\it \bar{h}}\,{\it \bar{q}}+{\it \bar{q}}+h\,{\it \bar{h}}\,q+q-h\,{\it \bar{h}}-1
 }\over{\sqrt{2}\,\sqrt{q+1}\,\sqrt{{\it \bar{q}}+1}\,\sqrt{q\,{\it \bar{q}}+h
 \,{\it \bar{h}}+1}\,\left(q\,{\it \bar{q}}-{\it \bar{q}}-q+2\,h\,{\it \bar{h}}+1\right)
 }}\;,
\ee

\be
e={{{\it \bar{h}}\,\left(q-1\right)}\over{q\,{\it \bar{q}}-{\it \bar{q}}-q+2\,h\,
 {\it \bar{h}}+1}}-{{\sqrt{2}\,{\it \bar{h}}\,\sqrt{q+1}\,\left(q-h\,{\it \bar{h}}-1
 \right)}\over{\sqrt{{\it \bar{q}}+1}\,\sqrt{q\,{\it \bar{q}}+h\,{\it \bar{h}}+1}\,
 \left(q\,{\it \bar{q}}-{\it \bar{q}}-q+2\,h\,{\it \bar{h}}+1\right)}}\;,
\ee

\be
f={{{\it \bar{h}}\,\left(q-1\right)}\over{q\,{\it \bar{q}}-{\it \bar{q}}-q+2\,h\,
 {\it \bar{h}}+1}}-{{\sqrt{2}\,{\it \bar{h}}\,\sqrt{q+1}\,\left(q\,{\it \bar{q}}-
 {\it \bar{q}}+h\,{\it \bar{h}}\right)}\over{\sqrt{{\it \bar{q}}+1}\,\sqrt{q\,
 {\it \bar{q}}+h\,{\it \bar{h}}+1}\,\left(q\,{\it \bar{q}}-{\it \bar{q}}-q+2\,h\,{\it \bar{h}}
 +1\right)}}\;,
\ee

\be
g={{h\,\left({\it \bar{q}}-1\right)}\over{q\,{\it \bar{q}}-{\it \bar{q}}-q+2\,h\,
 {\it \bar{h}}+1}}-{{\sqrt{2}\,h\,\sqrt{{\it \bar{q}}+1}\,\left(q\,{\it \bar{q}}-q+h
 \,{\it \bar{h}}\right)}\over{\sqrt{q+1}\,\sqrt{q\,{\it \bar{q}}+h\,{\it \bar{h}}+1}
 \,\left(q\,{\it \bar{q}}-{\it \bar{q}}-q+2\,h\,{\it \bar{h}}+1\right)}}\;,
\ee

\be
i={{h\,{\it \bar{h}}}\over{q\,{\it \bar{q}}-{\it \bar{q}}-q+2\,h\,{\it \bar{h}}+1}}+{{q^2
 \,{\it \bar{q}}^2+q\,{\it \bar{q}}^2-2\,{\it \bar{q}}^2-q^2\,{\it \bar{q}}+h\,{\it \bar{h}}\,
 q\,{\it \bar{q}}+3\,h\,{\it \bar{h}}\,{\it \bar{q}}+{\it \bar{q}}-h\,{\it \bar{h}}\,q-q+h\,
 {\it \bar{h}}+1}\over{\sqrt{2}\,\sqrt{q+1}\,\sqrt{{\it \bar{q}}+1}\,\sqrt{q\,
 {\it \bar{q}}+h\,{\it \bar{h}}+1}\,\left(q\,{\it \bar{q}}-{\it \bar{q}}-q+2\,h\,{\it \bar{h}}
 +1\right)}}\;,
\ee

\be \label{jparam}
j={{h\,\left({\it \bar{q}}-1\right)}\over{q\,{\it \bar{q}}-{\it \bar{q}}-q+2\,h\,
 {\it \bar{h}}+1}}-{{\sqrt{2}\,h\,\sqrt{{\it \bar{q}}+1}\,\left({\it \bar{q}}-h\,
 {\it \bar{h}}-1\right)}\over{\sqrt{q+1}\,\sqrt{q\,{\it \bar{q}}+h\,{\it \bar{h}}+1}
 \,\left(q\,{\it \bar{q}}-{\it \bar{q}}-q+2\,h\,{\it \bar{h}}+1\right)}}\;.
\ee
\normalsize
This satisfies all the requirements of a factorising twist, in particular $R=F_{21}F_{12}^{-1}$, 
as well as the triangular relation 
\be \label{twisttriang}
F_{21} F_{12}=I\otimes I \qquad \left(\text{in indices:}\quad  F^{i\;j}_{\;k\;l}F^{l\;k}_{\;n\;m}=\delta^i_{\;m}\delta^j_{\;n}\right)\;.
\ee
This relation allows us to express the $R$-matrix purely in terms of the twist itself, as
\be \label{Rfactored}
R_{12}=\left(F_{21}\right)^2 \qquad \left(\text{in indices:} \quad R^{i\;j}_{\;k\;l}=F^{j\;i}_{\;n\;m} F^{n\,m}_{\;\;l\;\;k}\right)\;,
\ee
a relation which will be used extensively later on. The twist is also of real type, satisfying the equivalent of (\ref{Rrealtype}). Together, these properties imply that $F$ is unitary as a $9\times 9$ matrix, a property which will also be very useful in what follows. 

By taking special cases of the $(q,h)$-twist we can easily reproduce previously known twists, such
as that for real $\beta$ \cite{Dlamini:2016aaa} or imaginary $\beta$ as above.\footnote{On the other hand, specialising to
  the  $w$-deformation we do not obtain the twist constructed in \cite{Dlamini:2016aaa}, which did not satisfy the triangularity
condition. We will comment on this difference later on.}

Of course, since the $(q,h)$ deformation depends on just four real parameters, the parameters (\ref{aparam})-(\ref{jparam})
are not all independent and there are numerous relations between them. We have:
\be
a=1-d-c \;,\; e=\hb (b-c)+f \;,\;j=\bar{e}=h(\bar{b}-\bar{c})+\bar{f} \;,\;i=\bar{b}\;,\; g=\bar{f}.
\ee
We also note that the parameters $c$ and $d$ are real, with
\be
d=c+\frac{\sqrt{2}(1-q\qb)}{\sqrt{1+q}\sqrt{1+\qb}\sqrt{1+q\qb+h\hb}}\;,
\ee
and that $b$ is simply related to $c$ by
\be
b=c+\frac{\sqrt{2}\sqrt{q+1}}{\sqrt{\qb+1}\sqrt{1+q\qb+h\hb}}\;.
\ee
These relations allow one to choose a convenient subset of variables in terms
of which to express the twist. Below we will use the (non-minimal and possibly non-optimal)
subset $\{b,\bb,c,f,\fb\}$.

\subsection{Exponential form of the twist}

In order to further study the $(q,h)$-twist, it is necessary to express it in
exponential form, i.e. as the exponentiation of a classical twist. The main reason is that
in order to construct the coassociator and the other structures of the Drinfeld-twisted
algebra we will need to act on the twist with coproducts, and it is not immediately clear how to evaluate
an expression such as $(\Delta\otimes\id)(F)$ using the group-like form (\ref{qhtwist}).
But since in  our case the undeformed coproduct is the
Lie-algebraic one, acting on Lie algebra elements and the identity as
\be \label{coproduct}
\Delta(X)=X\otimes I + I\otimes X \;,\quad \quad \Delta(I)=I\otimes I\;,
\ee
we can make use of the compatibility of the coproduct with multiplication to write 
\be
(\Delta\otimes \id)(F)=(\Delta\otimes \id)\left(e^{i f}\right)=(\Delta\otimes \id)\left(e^{i \sum f^{(1)}\otimes f^{(2)}}\right)
=e^{i\sum{\Delta(f^{(1)})\otimes f^{(2)}}}=e^{i(\Delta\otimes \id) f}\;.
\ee
So in order to compute $(\Delta\otimes \id)(F)$, we first compute $(\Delta\otimes \id)(f)$ (which is straightforward
since $f$ can be expressed in terms of generators of the algebra, on which the coproduct acts as in (\ref{coproduct}))
and then matrix exponentiate it.\footnote{As we will see,
the latter step is technically challenging for our matrices, but can always be performed given enough computing
power.} Writing a twist in exponential form also makes the
  co-unitality condition $(\epsilon\otimes \id)F=(\id\otimes \epsilon)F=1$ obvious, as the co-unit also exponentiates
  and $\epsilon(X)=0$.

By explicit calculation, we find that the full $(q,h)$ twist (\ref{qhtwist})  can be constructed by
suitably exponentiating a linear combination of the  four classical
twists corresponding to real and imaginary $\beta$, and real and imaginary $h$. We
will work with the usual Gell-Mann basis for the $\SU(3)$ algebra:
\be
\begin{split}
\lambda^1&=\begin{pmatrix}0&1&0\\1&0&0\\0&0&0\end{pmatrix},\;
\lambda^2=\begin{pmatrix}0&-i&0\\i&0&0\\0&0&0\end{pmatrix},\;
\lambda^3=\begin{pmatrix}1&0&0\\0&-1&0\\0&0&0\end{pmatrix},\;
\lambda^4=\begin{pmatrix}0&0&1\\0&0&0\\1&0&0\end{pmatrix},\;
\\
\lambda^5&=\begin{pmatrix}0&0&-i\\0&0&0\\i&0&0\end{pmatrix},\;
\lambda^6=\begin{pmatrix}0&0&0\\0&0&1\\0&1&0\end{pmatrix},\;
\lambda^7=\begin{pmatrix}0&0&0\\0&0&-i\\0&i&0\end{pmatrix},\;
\lambda^8=\frac1{\sqrt{3}}\begin{pmatrix}1&0&0\\0&1&0\\0&0&-2\end{pmatrix}\;.
\end{split}
\ee
Including also the identity as $\lambda^0$, the tensor products $\lambda^i\otimes \lambda^j$ of these matrices
form a basis for $\Acal\otimes \Acal$. 

To find the classical twists we need, we specialise the $(q,h)$-twist (\ref{qhtwist}) to each of the four
cases and expand to first order. Via a similar procedure to
the imaginary-$\beta$ case above, we can then express the full twist in each case as an exponentiation of the
corresponding classical twist:
\begin{itemize}
\item Real $\beta$ ($\qb=1/q,h=\hb=0$):
\be \label{realbtwist}
F_{\beta_r}=e^{\frac{i\beta}{2}\tilde{f}_{\beta_r}}=e^{\half\ln(q) \tilde{f}_{\beta_r}}=q^{\half\tilde{f}_{\beta_r}} \quad \text{with} \;\; \tilde{f}_{\beta_r}=\frac{\sqrt{3}}{2} \lambda^3\wedge \lambda^8
\ee

\item Imaginary $\beta$ ($\qb=q,h=\hb=0$):
\be
F_{\beta_i}=e^{i\arccos\left(\frac{1+q}{\sqrt{2}\sqrt{1+q^2}}\right) \tilde{f}_{\beta_i}}\quad \text{with} \;\; 
\tilde{f}_{\beta_i}=-\frac{1}{2}\left(\lambda^1\wedge\lambda^2-\lambda^4\wedge\lambda^5+\lambda^6\wedge \lambda^7\right)
\ee

\item Real $h$ ($q=\qb=1,h=\hb=h_r$):
  \be
  \begin{split}
F_{h_r}&=e^{i\frac{\sqrt{2}}2\text{arccos}\left(\frac{\sqrt{2}}{\sqrt{2+h_r^2}}\right) \tilde{f}_{h_r}} \quad \text{with} \;\;\\
  &\tilde{f}_{h_r}=\frac{1}{2}\left(\lambda^1\wedge \lambda^5+\lambda^7\wedge\lambda^1+\lambda^2\wedge\lambda^4+\lambda^2\wedge \lambda^6+\lambda^7\wedge \lambda^4+\lambda^6\wedge \lambda^5\right)
\end{split}
  \ee

\item Imaginary $h$ ($q=\qb=1,h=-\hb=i h_i$):
  \be \label{imhtwist}
  \begin{split}
  F_{h_i}&=e^{i\frac{\sqrt{2}}2 \text{arccos}\left(\frac{\sqrt{2}}{\sqrt{2+h_i^2}}\right) \tilde{f}_{h_i}}\quad \text{with}\;\;\\
 &   \tilde{f}_{h_i}=\half\left(\lambda^4\wedge\lambda^1+\lambda^1\wedge\lambda^6+\lambda^2\wedge \lambda^5+\lambda^2\wedge \lambda^7+\lambda^6\wedge\lambda^4+\lambda^5\wedge\lambda^7\right)
\end{split}
    \ee
\end{itemize}
where $x\wedge y=x\otimes y-y\otimes x$. 
We note that the $\tilde{f}$ classical twists are all hermitian as $9\times 9$ matrices (as expected, since $F$ is unitary). Adjusting
for our slightly different conventions, they
also individually satisfy the $r$-symmetric condition $r_{12}\sim f_{12}$ \cite{vanTongeren:2015uha}.\footnote{Expanding in terms
  of real parameters, we have $r_{\beta_r}=-\beta_r \tilde{f}_{\beta_r}$, $r_{\beta_i}=-(q-1)\tilde{f}_{\beta_i}$, $r_{h_r}=-h_r \tilde{f}_{h_r}$ and $r_{h_i}=-h_i \tilde{f}_{h_i}$. A relative factor of 1/2 arises on expanding the prefactors so one obtains $f_{12}=-\half r_{12}$ as required.} 
The abelian real-$\beta$ twist is well known \cite{Beisert:2005if,vanTongeren:2015uha,Dlamini:2016aaa}, while the other ones are new
(and clearly non-abelian).

It turns out that the exponential form of the general $(q,h)$ twist (\ref{qhtwist}) can be written in terms of the same classical
 twist matrices, but with coefficients which now each depend on all four parameters $(q,\qb,h,\hb)$. That is, we have:
\be
F_{q,h}=e^{i\, f_{q,h}}\;,
\ee
with the $(q,h)$ classical twist given by a linear combination of the four twists above
\be \label{classicaltwist}
f_{q,h}=\alpha_{\beta_r} \tilde{f}_{\beta_r}+ \alpha_{\beta_i} \tilde{f}_{\beta_i}+\alpha_{h_r} \tilde{f}_{h_r}
+\alpha_{h_i} \tilde{f}_{h_i}\;.
\ee
The expressions for the coefficients are:
\be
\begin{split}
\alpha_{\beta_r}&=\frac{i(b-\bb)(c-1) \rho}{\sqrt{1+b-c}\sqrt{1+\bb-c}\sqrt{(1-c)(3-b-\bb+c)-b\bb}}\;,\\
  \alpha_{\beta_i}&=\frac{(b \bb-(c-1)^2) \rho}{\sqrt{1+b-c}\sqrt{1+\bb-c}\sqrt{(1-c)(3-b-\bb+c)-b\bb}}\;,\\
    \alpha_{h_r}&=\frac{((1+b-c)f+(1+\bb-c)\fb) \rho}{\sqrt{1+b-c}\sqrt{1+\bb-c}\sqrt{(1-c)(3-b-\bb+c)-b\bb}}\;,\\
    \alpha_{h_i}&=\frac{i((1+b-c)f-(1+\bb-c)\fb) \rho}{\sqrt{1+b-c}\sqrt{1+\bb-c}\sqrt{(1-c)(3-b-\bb+c)-b\bb}}\;,
\end{split}
\ee
where
\be
\rho=\arccos\left[\half(b+\bb+\frac{f\fb}{c-1})\right]\;
\ee
and $c,\bb,f$ and $\fb$ are as defined in (\ref{aparam})-(\ref{jparam}). Note that all the $\alpha$ coefficients are real,
again as required for $F$ to be unitary.

These factors can be checked to reduce to the ones in (\ref{realbtwist})-(\ref{imhtwist}) in the corresponding limits.
Clearly, given the many relations between the parameters (\ref{aparam})-(\ref{jparam}), there are several alternative
ways to write the $\alpha$ coefficients, and it is quite possible that a more compact form for these can be found. The
choice of these four particular limits as our building blocks is of course arbitrary as well and other choices might
be considered.\footnote{As an example, we have found that switching from $\tilde{f}_{h_r}$ and $\tilde{f}_{h_i}$ to
  $\tilde{f}_{h_\pm}=\tilde{f}_{h_r}\pm i \tilde{f}_{h_i}$
  gives more compact expressions which simplify some explicit computations.}

This concludes our description of the general $(q,h)$ -twist, which is one of our main results. In the following
sections we will use this twist in order to study the quasi-Hopf algebraic structure of our theory and also to define a suitable star
product between the scalar fields of $\Ncal=4$ SYM which will lead to the general Leigh-Strassler theory.

\section{The coassociator}

In this section we construct the coassociator for the $(q,h)$-twisted algebra. The general expression for the twisted
coassociator is given in (\ref{coassociator}). However, in our case we are twisting the algebraic structure of the
$\Ncal=4$ SYM theory. So our starting point is the trivial coassociator $\Phi_0=I\otimes I \otimes I$ and the formula
simplifies to
\be \label{coassociatorspecific}
\Phi=F_{23}~(\id\otimes \Delta)(F) ~(\Delta\otimes \id)(F^{-1})~F_{12}^{-1}\;.
\ee
This expresses the coassociator as a product of four $3^3\times 3^3=27\times 27$ matrices, acting on triple copies of the
representation space of the algebra ordered as $\ket{111},\ket{112},\ldots$. Two of these matrices are trivially
defined as $F_{23}=I\otimes F$ and $F_{12}^{-1}=F^{-1}\otimes I$. So
the only non-trivial computations needed are the actions of $(\Delta \otimes \id)$ and  $(\id\otimes \Delta)$
on the twist. As discussed, to compute these expressions  we apply the coproducts to the
Taylor expansion of $e^{if}$ and re-exponentiate to find
\be
\begin{split}
(\Delta\otimes \id)(F)&=(\Delta\otimes \id)(e^{i\sum f_{(1)}\otimes f_{(2)}})=e^{i\sum \Delta(f_{(1)})\otimes f_{(2)}}\;,\\
(\id\otimes \Delta)(F)&=(\id\otimes \Delta)(e^{i\sum f_{(1)}\otimes f_{(2)}})=e^{i\sum f_{(1)}\otimes \Delta(f_{(2)})}\;,
\end{split}
  \ee
where we use Sweedler notation to write $f=\sum f_{(1)}\otimes f_{(2)}$ as shorthand for $f=\sum_{i,j}c_{ij} \lambda^i\otimes \lambda^j$.

Of course, the twisting procedure (\ref{coassociatorspecific}) guarantees that the coassociator we will obtain,
together with the $R$-matrix
(\ref{qhRmatrix}), will satisfy the quasi-Hopf Yang-Baxter equation (\ref{qhYBE}). So our main goal in explicitly
constructing the coassociator is not to check the qHYBE as such, but to confirm the validity of our computations
which will allow us to work with quasi-Hopf algebras in future applications.

\subsection{Imaginary $\beta$}

Let us again consider first the imaginary-$\beta$ case in order to illustrate the computation. Recall that the
twist (\ref{imbF}) can be expressed in terms of the $\su(3)$ generators as:
\be
F=e^{-i\frac{\alpha(q)}2 \left[\lambda^1\wedge \lambda^2-\lambda^4\wedge \lambda^5 +\lambda^6\wedge \lambda^7\right]}\;.
\ee
So we have, for instance:
\be
\begin{split}
(\id& \otimes \Delta) (F)=e^{-i\frac{\alpha(q)}2 \left[\lambda^1\otimes \Delta(\lambda^2)-\lambda^2\otimes \Delta(\lambda^1)
    -\lambda^4\otimes \Delta(\lambda^5)+\lambda^5\otimes \Delta(\lambda^4)
    +\lambda^6\otimes \Delta(\lambda^7)-\lambda^7\otimes \Delta(\lambda^6)\right]}\\
  &=\exp\big[ \!\!-\!i\frac{\alpha(q)}2 [\lambda^1\otimes(\lambda^2\otimes I+I\otimes \lambda^2)
      \!-\!\lambda^2\otimes (\lambda^1\otimes I+I\otimes \lambda^1)
      \!-\!\lambda^4\otimes(\lambda^5\otimes I +I \otimes \lambda^5)\\
  & \qquad\quad\; +\lambda^5\otimes (\lambda^4\otimes I +I \otimes \lambda^4)
      +\lambda^6\otimes (\lambda^7\otimes I +I \otimes \lambda^7)
      \!-\lambda^7\otimes (\lambda^6\otimes I +I\otimes \lambda^6)]\big]\;.
\end{split}
\ee
This matrix can be easily computed using computer algebra. Similarly we can
compute the expression $(\Delta \otimes \id)(F^{-1})$. Finally we multiply the four
matrices in (\ref{coassociatorspecific}). The final result is presented in appendix \ref{imbcoassociator},
where for clarity we do not give the matrix form of $\Phi$ but rather its tensorial
components expressed in the basis $|i\rangle\otimes |j\rangle\otimes  |k\rangle$.

\subsection{The general case} \label{generalcoassocsec}

After illustrating the steps involved in the computation for the special case of imaginary $\beta$,
we can now proceed to evaluate the coassociator for the general $(q,h)$-twist (\ref{qhtwist}). This turns
out to be considerably more computationally demanding, and we have not 
found it possible to exhibit the components of $\Phi$ in the same compact way as for the imaginary-beta case. We
will therefore outline the steps of the computation and the interested reader can find the details
in the Mathematica file associated to this preprint.

As above, the main elements in constructing the coassociator will be the two $27\times 27$ matrices
$(\Delta\otimes \id)(f)$ and $(\id\otimes \Delta)(f)$, where $f$ is the classical twist (\ref{classicaltwist}).
For concreteness, we write them down in appendix \ref{matrices}. Direct exponentiation of these matrices
in Mathematica appears challenging, so to exponentiate them we will need to explicitly find their diagonalisation
matrices $V$, which allow us to find the exponential of a matrix by way of its diagonalised version: 
\be
M=V^\dag D V \;\quad \Rightarrow \quad e^M=V^\dag e^D V \;.
\ee
We have chosen to construct the left-eigenvectors, such that $E^{(L)}_i (\Delta\otimes \id)(f)=\lambda_i E^{(L)}_i$.\footnote{Note that the
  $(L)$ refers to the coproduct being on the left in  $(\Delta\otimes \id)(f)$, and not to the type of eigenvector.} Then
$V^{(L)}$ is the matrix whose rows are the eigenvectors. 
Noting that $(\Delta\otimes \id)(f)$ and $(\id \otimes \Delta)(f)$ are related by conjugation with $P_{13}$, they have
the same eigenvalues, and their eigenvectors are related as $E^{(R)}_i=E^{(L)}_{i'} P_{13}$, where the $i'$ means that some
indices need to be swapped in $E_i^{(R)}$ in order to keep the eigenvalues in the same order. The diagonalisation
matrix $V^{(R)}$ can then be constructed from these eigenvectors. Finally, we can write
\be \label{DidFandidDFexp}
\begin{split}
(\Delta\otimes \id)(F)&=e^{i(\Delta\otimes \id)(f)}=(V^{(L)})^\dag e^D V^{(L)}\;,\\
(\id\otimes \Delta)(F)&=e^{i(\id\otimes \Delta)(f)}=(V^{(R)})^\dag e^D V^{(R)}\;,
\end{split}
  \ee
where $D$ is the diagonal matrix given by $D_{ii}=e^{\lambda_i}$. We similarly compute
\be
\begin{split}
(\Delta\otimes \id)(F^{-1})&=e^{-i(\Delta\otimes \id)(f)}=(V^{(L)})^\dag e^{-D} V^{(L)}\;,\\
(\id\otimes \Delta)(F^{-1})&=e^{-i(\id\otimes \Delta)(f)}=(V^{(R)})^\dag e^{-D} V^{(R)}\;.
\end{split}
  \ee
Using these components, we can construct $\Phi$ and its inverse according to (\ref{coassociatorspecific}),
and are finally able to verify the quasi-Hopf YBE (\ref{qhYBE}). This check is performed through choosing generic numerical
values for the parameters $q,\qb,h,\hb$, and is successful. The details of the above construction and the
verification of the qHYBE can be found in the Mathematica file submitted together with this article\footnote{If one is
  interested in a specific choice of parameters it is possible, with some patience, to verify the qHYBE analytically. In the
  associated Mathematica file, we illustrate this for the cases of the imaginary-$\beta$ and  real-$h$ deformations.}.

One possible issue with our construction of the coassociator is that it relies on the exponentiations (\ref{DidFandidDFexp}).
This is certainly fine for values of the parameters close to their classical values, but one should perhaps
take additional care when considering, for instance, roots of unity, or very non-classical limits such as $q\rightarrow 0, \hb=1/h$
(a case which is interesting both from the perspective of planar finiteness \cite{Bork2008} as well as integrability
\cite{Bundzik:2005zg, Mansson:2008xv}), or the ``cubic'' model of  \cite{Minahan:2011dd}.
Even for some special cases which are close to classical values, such as the  $w$-deformation
$q=\qb=1+w,h=\hb=w$ \cite{Dlamini:2016aaa}, we have
noticed that some denominators in the diagonalisation matrices become large and the numerical evaluation of
the coassociator cannot be trusted. Fortunately, in this case it is possible to simplify the matrices analytically
and confirm that the coassociator becomes trivial, as it should since the $R$-matrix satisfies the usual YBE. 

As mentioned, the qHYBE is guaranteed to be satisfied by the twisting procedure, so confirming it
is really a check of our computation of the coassociator and the ingredients that went into it. This is important,
if one would like to perform explicit computations in the quasi-Hopf algebra in the future. (Some such
computations appear in the next section).

Another ingredient in confirming the quasi-Hopf structure is showing the pentagon identity (\ref{pentagon}).
In appendix \ref{pentagonapp} we discuss how this identity follows from the twisting procedure. The
antipode and elements $\alpha$ and $\beta$ of the quasi-Hopf algebra are also guaranteed by the twisting
procedure, and since we have not needed their explicit forms, we have not computed them explicitly.

We conclude that, as expected, the twist (\ref{qhtwist}) acting on the trivial $R$-matrix $R=I\otimes I$ has led to
a triangular quasi-Hopf algebra. Since this algebra is a Drinfeld twist of the universal enveloping algebra of the
$\su(3)$ Lie-algebraic symmetry of $\Ncal=4$
SYM, we can denote it as $U_{q,h}[\su(3)]$, where $(q,h)$ can also be taken to stand for ``quasi-Hopf''.\footnote{One
  could also use the notation $\SU(3)_{q,h}$, but it might be best to reserve that for  the dual FRT picture studied
  in this context in \cite{Mansson:2008xv}. In using this notation one should keep in mind that this algebra is
  unrelated to multiparameter deformations of $\SU(3)$, such as those in e.g. \cite{Reshetikhin90,EwenOgievetsky94},
  which are Hopf and not quasi-Hopf.}

\section{The  star product}

After establishing the quasi-Hopf structure of our $(q,h)$-twisted algebra, let us now
turn to  its module, or representation space, focusing on the fundamental
representation. As discussed in e.g. \cite{Majid,Grosse:2001pr,Blohmann03,Szabo:2006wx}, when twisting
an initial Hopf algebra, and thus defining a new coproduct according to
(\ref{twistedcoproduct}), compatibility between the algebra coproduct $\Delta_F$ and the module product $m$
requires the latter to be twisted as well. The new product is
\be
m_F(a\otimes b) =m (F^{-1}\triangleright x\otimes y) =\sum (F^{-1}_{(1)}\triangleright a) \cdot (F^{-1}_{(2)}\triangleright b)\;.
\ee
where we abuse notation by also denoting $m$ by $\cdot$, which is also used for the product in $\Acal$. 
It is natural to express this modified module product as a star product:
\be \label{starprod}
a\star b = m_F(a\otimes b)\;.
\ee
For explicit computations, it is useful to write this and similar expressions in
index notation. For this we define a basis $e_i$ for our vector space
\be
e_1=\triplet{1}{0}{0}\;,\quad e_2=\triplet{0}{1}{0}\;,\quad e_3=\triplet{0}{0}{1}\;,
\ee
and notice that the star product (\ref{starprod}) is essentially a rotation of the vector $a\otimes b=c^{ij}e_i\otimes e_j$
by the action of $F^{-1}$. In terms of the $e_i$, this action can be written as $(e_i\star e_j)=e_k e_l (F^{-1})^{k\;l}_{\;i\;j}$.
However, since we would like to express the star product as a transformation on the coordinates, we will be interested
in the dual basis $z^i\otimes z^j$, related to the basis $e_k\otimes e_l$ by
$\langle z^i\otimes z^j,e_k\otimes e_l\rangle=\delta^i_k\delta^j_l$. In order for the same relation to hold for the $\star$-deformed
dual basis, we need to transform it as $z^i\star z^j=F^{i\;j}_{\;k\;l}z^k\otimes z^l$ (where we used $F_{21}=F^{-1}_{12}$). The outcome of this is
that the star product acting on the coordinates $z^i$ is given by
\be \label{starindices}
z^i\star z^j= F^{i\;j}_{\;k\;l} z^k z^l\;.
\ee
In \cite{Dlamini:2016aaa}, this star product was used to relate the related the superpotentials
of the $\beta$ and $w$-deformed theories to that of the $\Ncal=4$ SYM theory. Since we now have
the full $(q,h)$-twist at our disposal, we simply extend the star product to the
general case. We note that the $z^i$ on the right-hand-side of (\ref{starindices}) are in principle commuting
  coordinates, but in the field theory context they will acquire
additional non-abelian structure (as they will be mapped to matrix-valued scalar fields) so we
will be careful to preserve their ordering unless otherwise noted.

Of course, an essential difference between \cite{Dlamini:2016aaa} and the current work is that the twist
(\ref{starprod})  does not satisfy the YBE. Working out the two possible cubic products, one finds the expressions:
\be \label{cubicstars}
\begin{split}
a\star(b\star c)=m\left(((\id\otimes \Delta)(F^{-1}) (1\otimes F^{-1}))\triangleright [a\otimes b \otimes c]\right)\;,\\
(a\star b) \star c =m\left(((\Delta\otimes \id)(F^{-1}) (F^{-1}\otimes 1))\triangleright [a\otimes b \otimes c]\right)\;. 
\end{split}
\ee
These will be different in general. Note that in \cite{Dlamini:2016aaa} we made use of the abelian nature
of the twists appearing there to rewrite the above expressions as cubic products of $F^{-1}$. However, the $(q,h)$-twist is
non-abelian, so that option is not available here. In order to compute these two
cubic products we need to first compute the explicit $27\times 27$ matrices in (\ref{cubicstars}).
As before, we will first show how to do this for the imaginary-$\beta$ deformation before proceeding to the
general case. 

\subsection{Imaginary $\beta$}

To illustrate the computations without the expressions becoming too unwieldy,
let us again specialise first to the imaginary $\beta$ deformation. In order to compute
the expressions in (\ref{cubicstars}), we need to apply the coproduct on the twist $F$. As discussed above,
the coproduct
$\Delta$ entering here is that of the undeformed theory, so its action on generators of the algebra
is the trivial one (\ref{Liecoproduct}).
So in order to apply say $\id\otimes \Delta$ on the twist we need to first write $F$ in terms of generators
of $\SU(3)$. We recall that $F_{12}=e^{if_{12}}$, where in this case $f_{12}$ is
\be
f_{12}=-\frac{\alpha(q)}2 \left(\lambda_1\wedge \lambda_2-\lambda_4\wedge \lambda_5 +\lambda_6\wedge \lambda_7\right)\;,
\ee
with $\alpha(q)$ as in (\ref{alphaq}). So we compute:
\be \label{xyzstarL}
(a\star b)\star c=m\left((\Delta\otimes \id)(e^{-if}) F^{-1}_{12}\triangleright [a\otimes b \otimes c]\right)
=m\left(e^{-i\sum\Delta(f_{(1)}) \otimes f_{(2)}} F^{-1}_{12}\triangleright [a\otimes b \otimes c]\right),
\ee
while 
\be \label{xyzstarR}
a\star(b\star c)=m\left((\id\otimes \Delta)(e^{-if}) F^{-1}_{23}\triangleright [a\otimes b \otimes c]\right)
=m\left(e^{-i\sum f_{(1)} \otimes \Delta(f_{(2)})} F^{-1}_{23}\triangleright [x\otimes y \otimes z]\right).
\ee
For ease of reference, let us give names to the $27\times 27$ matrices appearing in the cubic star products,
according to whether the parenthesis is placed to the left or to the right:
\be \label{cubicFs}
[F_{3,L}]=(\Delta\otimes \id)(F^{-1}) (F^{-1}\otimes 1) \;\;\;\text{and}\;\; [F_{3,R}]=(\id\otimes \Delta)(F^{-1}) (1\otimes F^{-1})\;.
\ee
The explicit expressions for these matrices are given in appendix \ref{imbcubic}, where (as for the coassociator)
we find it easier to record their components as tensors acting on three copies of the algebra. In general, these
are unitary matrices, although in the imaginary-$\beta$ case they are orthogonal.

As in the quadratic case, we would also like to express the cubic star product in terms of its action on
our coordinate basis $z^i$. This will again entail an inversion of the matrix, so we write:
\be \label{zstarindices}
(z^i\star z^j)\star z^k=[F^{-1}_{3,L}]^{i\;j\;k}_{\;i'j'k'} z^{i'} z^{j'} z^{k'}\;\;\;\text{and}\;\;
z^i\star (z^j\star z^k)=[F^{-1}_{3,R}]^{i\;j\;k}_{\;i'j'k'} z^{i'} z^{j'} z^{k'}\;.
\ee
where
\be \label{cubicFinvs}
[F^{-1}_{3,L}]= (F\otimes 1) (\Delta\otimes \id)(F)\;\;\;\;\text{and}\;\; 
[F^{-1}_{3,R}]= (1\otimes F) (\id\otimes \Delta)(F)\;.
\ee
As the matrices tabulated in appendix \ref{imbcubic} are orthogonal, their inverses can easily be found by transposition,
so we do not record them here. Of course, by construction, the left- and right- matrices are related:
\be \label{LRcubic}
[F_{3,L}]=[F_{3,R}]\Phi \;\quad \text{and} \quad [F_{3,L}^{-1}]=\Phi^{-1}[F_{3,R}^{-1}]\;,
\ee
which shows explicitly how the coassociator relates the two cubic bracketings.

\subsection{The $(q,h)$-star product} \label{qhstarsec}

We can now proceed to define the $(q,h)$ star product using the definition (\ref{starindices}) with the twist (\ref{qhtwist}). 
To see why this is a consistent definition, it important to recall the link to quantum planes, which in this context were
discussed in detail in \cite{Mansson:2008xv}. In the context of the FRT relations defining a Hopf
algebra (which is a dual picture to the UEA picture that we are using here), the quantum plane can directly be seen as
the space on which the algebra acts, and its non-commutativity is governed by the same $R$-matrix that
enters the FRT relations through expressions such as $R{\bf x_1}{\bf x_2}={\bf x_2}{\bf x_1}$.\footnote{Quantum planes acted
  on by matrix quantum groups which are deformations of $\mathrm{GL}(3)$, and the corresponding $R$-matrices, were
  classified in \cite{EwenOgievetsky94}, however those planes were alphabetically ordered, rather than the cyclically ordered
  planes which arise in our context.}

From a physical perspective, the
quantum plane relations are identified with the F-term relations of the gauge theory, as described in \cite{Berenstein:2000ux}.
To see this, it is useful to have in mind the usual setup where a stack of $N$ D3-branes is placed at the origin of a transverse
$\Cset_3$. On the worldvolume of the branes, these six transverse directions appear as scalar fields in the adjoint
representation of $\SU(N)$ \cite{Witten:1995im}. Since the coordinates of $\Cset_3$ all commute, all these directions are
equivalent, and the action of the scalars will have an $\SO(6)$ symmetry (though of course only an $\SU(3)\times \Urm(1)$ is
explicit if we insist on working with holomorphic coordinates). This symmetry is also reflected on the moduli space of the
worldvolume theory, resulting in the usual F-term relations of $\Ncal=4$ SYM of the form $[\Phi^i,\Phi^j]=0$. If we now replace
the transverse space by a quantum plane $\Cset_3^{q,h}$, the worldvolume action will reflect the $U_{q,h}[\su(3)]$ symmetry of this
space and translate to quantum group relations for the scalar fields, of the type $[\Phi^i,\Phi^j]_{q,h}=0$. These were the
relations discussed in \cite{Berenstein:2000ux}. Given the above, in the following we will often switch between the quantum plane geometry
of the transverse space and relations for the $\Ncal=4$ SYM scalars. 

In \cite{Mansson:2008xv}, the quantum plane coordinates were non-commutative, while in the present (dual) setting
the non-commutativity is controlled by the star product. So we expect a star-product version of the quantum plane relations in
\cite{Mansson:2008xv} to hold:
\be \label{holqplane}
R^{i\;j}_{\;k\;l} z^k \star z^l=z^j\star z^i\;.
\ee
Noting that (\ref{starindices}), together with (\ref{twisttriang})  implies
\be
z^i\star z^j=F^{i\;j}_{\;k\;l} z^k z^l \;\;\Rightarrow \;\; z^mz^n=F^{n\;m}_{\;j\;i} z^i\star z^j\;,
\ee
we compute
\be
z^j\star z^i=F^{j\;i}_{\;n\;m} z^n z^m = F^{j\;i}_{\;n\;m} z^m z^n=F^{j\;i}_{\;n\;m} \left(F^{n\,m}_{\;l\;\;k} z^k\star z^l\right)
=F^{j\;i}_{\;n\;m} F^{n\,m}_{\;l\;\;k} z^k\star z^l=R^{i\;j}_{\;k\;l}z^k\star z^l,
\ee
where we used (\ref{Rfactored}) in the last step. This is just as required from (\ref{holqplane}), so our star
product is consistent with the underlying quantum plane structure.\footnote{Note that in the above derivation we
  made use of the undeformed (commutative) quantum plane relation, $z^mz^n=z^nz^m$. In the gauge theory context, this
  corresponds to imposing the $\Ncal=4$ SYM F-term relations $[\Phi^m,\Phi^n]=0$.  However, the star product relates
  LS and $\Ncal=4$ SYM expressions without necessarily imposing this constraint.} Let us note that a similar star-product
derivation of quantum plane relations appears in \cite{Grosse:2001pr}, and a detailed study of the interplay
between quantum symmetries, quantum planes and star products can be found in \cite{Blohmann03}.

We can thus expect that expressions which are natural from the quantum plane picture will also be natural from the
perspective of the star product, and this indeed turns out to be the case. For instance, as a special case of (\ref{holqplane}),
we find:
\be \label{qplanestar}
z^1\star z^2-q z^2\star z^1+h z^3\star z^3=\frac{\sqrt{1+q}\sqrt{1+q\qb+h\hb}}{\sqrt{2}\sqrt{1+\qb}}\left(z^1z^2-z^2z^1\right)\;,
\ee
as well as its cyclic permutations. On the right hand side the $z$'s are multiplied using the undeformed module product. As
this product is commutative, the right hand side vanishes, demonstrating
how the star product reproduces the quantum plane relations appropriate to the general
Leigh-Strassler theory \cite{Mansson:2008xv}.

As discussed, in the gauge theory picture the $\Cset^3$ coordinates $z^i$ are replaced by the adjoint scalar fields $\Phi^i$.
Being matrices, these scalars do not generically commute, so the relation above should be read as
\be
\Phi^1\star \Phi^2-q \Phi^2\star \Phi^1+h~ \Phi^3\star \Phi^3=\frac{\sqrt{1+q}\sqrt{1+q\qb+h\hb}}{\sqrt{2}\sqrt{1+\qb}}~[\Phi^1,\Phi^2]
\ee
plus cyclic permutations. This illustrates how the $F$-term constraints in the Leigh-Strassler-deformed theories (discussed from a quantum
plane perspective in \cite{Berenstein:2000ux}) can be simply related to the $\Ncal=4$ SYM $F$-term constraints using the $\star$-product. 
Clearly, both sides reduce to the standard commutator in the classical limit $(q,h)=(1,0)$.

Applying the star product to cubic expressions, by defining the analogues of (\ref{cubicFs}),
it is easy to establish that it is non-associative for generic $q,h$. 
However, by construction, the coassociator allows us to relate any two placements of parentheses. For the star
product on vectors the relation is simply:
\be
(a\star b)\star c=\sum [\Phi_{(1)}\trr a]\star \left([\Phi_{(2)}\trr b] \star [\Phi_{(3)} \trr c]\right)\;.
\ee
This is a standard result (see e.g.\cite{Grosse:2001pr,Mylonas:2013jha}), however we would also like to express it
in index form. Rewriting (\ref{LRcubic}) in tensor language and recalling the definition (\ref{zstarindices}), we can write:
\be \label{LRrelationstar}
(z^i\star z^j)\star z^k=\Phi^{i\;j\;k}_{\;l\;m\;n}\, z^l\star(z^m\star z^n)\;,
\ee
an equality which can be verified by explicitly evaluating both sides. This formula allows us to
convert any left-bracketed cubic expression to a sum of right-bracketed ones. (And vice-versa,
using the inverse coassociator).

A very interesting relation arises when we introduce the star product into the Leigh-Strassler superpotential itself.
By explicit computation we find:
\small
\be
\begin{split}
&(z^1\star z^2)\star z^3\!+\!(z^2\star z^3)\star z^1\!+\!(z^3\star z^1)\star z^2
\!-\!q((z^1\star z^3)\star z^2\!+\!(z^2\star z^1)\star z^3\!+\!(z^3\star z^2)\star z^1)\\
&\qquad\!+\!h ((z^1\star z^1)\star z^1\!+\!(z^2\star z^2)\star z^2\!+\!(z^3\star z^3)\star z^3)\\
&\quad=\frac{\sqrt{1\!+\!q}\sqrt{1\!+\!q\qb\!+\!h\hb}}{\sqrt{2}\sqrt{1\!+\!\qb}}\left((z^1 z^2) z^3\!+\!(z^2 z^3) z^1\!+\!(z^3 z^1) z^2
\!-\!(z^1 z^3) z^2\!-\!(z^2 z^1) z^3\!-\!(z^3 z^2) z^1\right)
\end{split}
\ee
\normalsize
Switching to the right bracketing in the above expression produces the same answer. 
Replacing $z^i\rightarrow \Phi^i$ and taking the gauge theory trace, we find that not only the quadratic quantum plane
relation (\ref{qplanestar}) but also the cubic Leigh-Strassler superpotential (\ref{LSW}) 
reduces to its corresponding undeformed ($\Ncal=4$ SYM) expression upon introducing the $\star$ -product.

\subsection{The inverse star product} \mbox{}

As discussed more extensively in \cite{Dlamini:2016aaa}, it is very convenient to also introduce the \emph{inverse} star product, which
we will denote by an asterisk.
\be
a\astv b = m_{F^{-1}}(x\otimes y)=m (F\triangleright a\otimes b) =\sum(F_{(1)}\triangleright a) \cdot (F_{(2)}\triangleright b)\;.
\ee
Similar arguments as in the previous section can be used to show that,
acting on the coordinates, the $\astv$ -product is simply:
\be \label{istarindices}
z^i\astv z^j=(F^{-1})^{i\;j}_{\;k\;l} z^k z^l=F^{j\;i}_{\;l\;k} z^kz^l\;.
\ee
Using the explicit form of the twist, we can check the following relation (plus its cyclic analogues):
\be
z^1\astv z^2-z^2\astv z^1=\frac{\sqrt{2}{\sqrt{1+\qb}}}{\sqrt{1+q}\sqrt{1+q\qb+h\hb}}\left(z^1z^2-q z^2z^1+h (z^3)^2\right)\;,
\ee
which is of course the inverse relation to (\ref{qplanestar}). In the gauge theory context, this relation
tells us that $\astv$ -deforming the commutator $[\Phi^1,\Phi^2]$ leads to the
corresponding $(q,h)$-commutator:
\be
  \Phi^1\astv \Phi^2-\Phi^2\astv\Phi^1=\frac{\sqrt{2}{\sqrt{1+\qb}}}{\sqrt{1+q}\sqrt{1+q\qb+h\hb}}\, [\Phi^1,\Phi^2]_{q,h}\;,
   \ee
where we define $[\Phi^1,\Phi^2]_{q,h}=\Phi^1\Phi^2-q\,\Phi^2\Phi^1+h(\Phi^3)^2$.  In other words,
this product takes $\Ncal=4$ SYM expressions to ones relevant to the Leigh-Strassler theory. 

At cubic level, the two bracketings of the $\astv$ -product are:
\be
(a\astv b)\astv c=m\big((F\otimes 1)(\Delta\otimes \id)(F)\triangleright [a\otimes b \otimes c]\big)
\ee
and
\be
a\astv(b\astv c)=m\big((1\otimes F) (\id\otimes \Delta)(F)\triangleright [a\otimes b \otimes c]\big),
\ee
where the relevant $27\times 27$ matrices appear in (\ref{cubicFinvs}).

As before, for explicit computations we would like to know how the $\astv$ -product acts on the coordinates themselves.
Given the inversion that occurs when switching to coordinates, we find
\be \label{zcubicastL}
(z^i\astv z^j)\astv z^k=[(\Delta\otimes \id)(F^{-1})(F^{-1}\otimes 1)]^{ijk}_{i'j'k'} z^{i'}z^{j'}z^{k'}=[F_{3,L}]^{ijk}_{i'k'j'} z^{i'} z^{j'} z^{k'}\;,
\ee
and similarly:
\be \label{zcubicastR}
z^i\astv (z^j\astv z^k)=[(\id\otimes \Delta)(F^{-1})(1\otimes F^{-1})]^{ijk}_{i'j'k'} z^{i'}z^{j'} z^{k'}=[F_{3,R}]^{i\;j\;k}_{\;i'j'k'} z^{i'} z^{j'} z^{k'}\;.
\ee
For the imaginary-$\beta$ case, the $[F_{3,L}]$ and $[F_{3,R}]$ matrices are given in appendix \ref{imbcubic}, while
for the general $(q,h)$ case they can be computed using the Mathematica file associated to this preprint.

We now have all the information we require in order to compute cubic star-product expressions.
 As expected, the cubic star product is not associative. As an example, for imaginary-$\beta$ we compute:
\small
\be
(z^1\astv z^2)\astv z^2=\frac{1+q\cos(\sqrt{2}\alpha)}{\sqrt{2}\sqrt{1+q^2}} z^1z^2z^2+\frac{\cos(\sqrt{2}\alpha)-1}{\sqrt{2}\sqrt{1+q^2}} z^2z^1z^2-\frac{\sin(\sqrt{2}\alpha)}{\sqrt{2}} z^2z^2z^1\;,
 \ee
while:
 \be
 z^1\astv(z^2\astv z^2)=\cos(\sqrt{2}\alpha) z^1z^2z^2-\frac{q\sin(\sqrt{2}\alpha)}{\sqrt{1+q^2}}z^2z^1z^2-\frac{\sin(\sqrt{2}\alpha)}{\sqrt{1+q^2}}z^2z^2z^1\; .
 \ee
 \normalsize
 However, as for the $\star$ -product, we can relate any left-bracketed expression to a sum
 of right-bracketed ones using the coassociator. Since in this case the coassociator appears on the
 right in (\ref{LRcubic}), the relation cannot be written as simply as (\ref{LRrelationstar}), but can still
 be expressed compactly in a hybrid of Sweedler and index notation as
 \be
 (z^i\astv z^j)\astv z^k=\sum \Phi_{(1)}{}^i_{\;l}z^l \astv (\Phi_{(2)}{}^j_{\;m} z^m\astv \Phi_{(3)}{}^k_{\;n} z^n)\;.
 \ee
 Again, the equality of the two sides can be explicitly verified.
 
The inverse star product allows us to obtain relations in the generic Leigh-Strassler-deformed
theory by star-deforming $\Ncal=4$ SYM expressions, with the (non-trivial) caveat that it is non-associative and
thus one can only expect to obtain correct $(q,h)$-deformed expressions if one knows how the parentheses should
be placed in the undeformed ones, a situation similar to the one encountered when deforming from a commutative
setting to a non-commutative one. 

\subsection{The cyclic star product} \label{cyclicstar}

As we saw, the star products defined in the previous sections are non-associative. So, if for instance
we wish to use the $\astv$ -product to deform an $\Ncal=4$
SYM expression involving the fields  to the $(q,h)$ theory, the result will depend on how the parentheses
are placed in the undeformed expression and can result in ambiguous expressions. 

However, and rather surprisingly, it turns out that this issue 
is not present when considering precisely the (cyclically symmetrised) cubic
expression which appears in the superpotential of the theory. In that case, the two different placements of
parentheses are equal. So for instance one can show, for the general $(q,h)$-twist, that
\be
z^1\astv(z^2\astv z^3)+z^2\astv(z^3\astv z^1)+z^3\astv(z^1\astv z^2)= (z^1\astv z^2) \astv z^3+(z^2\astv z^3) \astv z^1+(z^3\astv z^1) \astv z^2
\ee
and similarly for the expression with any two indices interchanged. 
Therefore, if we apply the cyclicity of the trace to write the $\Ncal=4$ SYM superpotential as 
\be
\frac{1}{3}\Tr\left[\Phi^1\Phi^2\Phi^3+\Phi^2\Phi^3\Phi^1+\Phi^3\Phi^1\Phi^2-
\Phi^1\Phi^3\Phi^2-\Phi^3\Phi^2\Phi^1-\Phi^2\Phi^1\Phi^3\right]\;,
\ee
there is no need to be careful with parentheses, as either of the two placements will work (as long as
the same choice is used for all the terms).\footnote{The need to rewrite the trace to make the cyclicity evident
  before inserting the star product was already seen in the associative setting of \cite{Dlamini:2016aaa}, and has
  also appeared in a similar context in \cite{Garus:2017bgl} (see also \cite{Beisert:2018zxs} for further discussion
  of the interplay between cyclicity and quantum groups). } Choosing the left placement, star-deforming and performing
the explicit computations, the above expression becomes 
\small
\be
\begin{split}
\frac{1}{3}&\Tr\left[(\Phi^1\astv\Phi^2)\astv\Phi^3\!+\!(\Phi^2\astv\Phi^3)\astv\Phi^1\!+\!(\Phi^3\astv\Phi^1)\astv\Phi^2\!-\!
(\Phi^1\astv\Phi^3)\astv\Phi^2\!-\!(\Phi^3\astv\Phi^2)\astv\Phi^1\!-\!(\Phi^2\astv\Phi^1)\astv\Phi^3\right]\\
&= \frac{\sqrt2 \sqrt{1+\qb}}{3\sqrt{1+q}\sqrt{1+h\hb+q\qb}}\Tr[\Phi^1 \Phi^2 \Phi^3+\Phi^2 \Phi^3 \Phi^1+\Phi^3 \Phi^1 \Phi^2
-q(\Phi^1 \Phi^3 \Phi^2+\Phi^3 \Phi^2 \Phi^1+\Phi^2 \Phi^1 \Phi^3)\\
&\qquad\qquad\qquad\qquad\qquad\qquad  +h(\Phi^1 \Phi^1 \Phi^1+\Phi^2 \Phi^2 \Phi^2+\Phi^3 \Phi^3 \Phi^3)]\;.
\end{split}
\ee
\normalsize
Here the product between the fields is just the usual matrix product, so the cyclically related terms are equal under the
trace. 
Introducing a more compact notation where it is understood that one needs to first write out cyclically related terms before
introducing the $\astv$ -product, we can write
\be \label{starW}
g\Tr\left[\Phi^1\astv \Phi^2\astv \Phi^3\!-\!\Phi^1\astv \Phi^3\astv \Phi^2\right]=\kappa ~\Tr\left[\Phi^1\Phi^2\Phi^3\!-\!q \Phi^1\Phi^3\Phi^2+
   \frac h3\left((\Phi^1)^3\!+\!(\Phi^2)^3\!+\!(\Phi^3)^3\right)\right]\;,
\ee
which is the Leigh-Strassler superpotential (\ref{LSW}) for the specific value 
\be
\kappa=g\frac{\sqrt2 \sqrt{1+\qb}}{\sqrt{1+q}\sqrt{1+h\hb+q\qb}}\;.
\ee
We have thus achieved the long-standing goal of obtaining the full Leigh-Strassler
superpotential from a star product inserted in the superpotential of $\Ncal=4$ SYM. This generalises
star products for associative cases in \cite{Lunin:2005jy,Bundzik2007,Dlamini:2016aaa}. We emphasise
that, even though the $\astv$ -product is non-associative in general, the coassociator does not appear
to be necessary for this specific result.

Substituting the value of $\kappa$ in the conformal constraint equation (\ref{conformalconstraint}), we
find that it is perfectly consistent with it in the planar limit (which makes sense, as we have only used planar
information in our construction). Also, we are here of course only dealing with the classical lagrangian, so we
would certainly not expect to see any signs of the corrections to the constraint equation that arise at higher
loop orders (in particular, at four loops at planar level \cite{Bork2008}). But one could hope that our star-product
approach will allow for more efficient computations of loop
corrections in the Leigh-Strassler theories, and through that a better understanding of the conformal condition
at higher loops.

Let us note that specifying the parameters to those of the $w$-deformation of \cite{Dlamini:2016aaa} one obtains
the correct superpotential, however the coefficient $\kappa$ in that work as different. This is to be expected,
since our twist here is different from the one used in \cite{Dlamini:2016aaa}. We interpret that twist (which
is non-equivalent to the present one, being non-triangular) as deforming
a given $\Ncal=4$ SYM theory (labelled by the value of the gauge coupling $g$) to a different superconformal
deformation which has the same values for $q$ and $h$ but differs also in the gauge coupling $g'$, and thus
also in the $\kappa$ parameter in front of the superpotential (\ref{LSW}). Whether this can be made precise by
understanding the twist of \cite{Dlamini:2016aaa} as a two-step process of first
changing the gauge coupling along the $\Ncal=4$ SYM marginal line and then performing a triangular twist is an interesting
open question.

We saw that the non-associativity of our quasi-Hopf-twisted algebra does not enter for the specific
combination of fields appearing in the superpotential. The underlying reason is not clear to us\footnote{At the practical level, of course, one can easily  see when taking differences of the two placements of parentheses that one obtains differences of terms where
  the $z^i$'s are permuted, so cyclically symmetrising them gives zero.} but it is likely to be related to another
curious fact: In \cite{Mansson:2008xv} associativity was imposed by hand on the cubic relations between generators
arising from the FRT relations, but still led to a consistent, central, quantum determinant. This was precisely as required
to show Hopf invariance of the superpotential. In the current article, we work in the universal enveloping algebra picture
(dual to the FRT picture employed in \cite{Mansson:2008xv}), so what we are seeing is perhaps the dual 
statement to what appeared there. In other words, although in \cite{Mansson:2008xv} associativity was imposed in
a situation where one did not expect it to hold, it was consistent with the cubic expression arising in the quantum
derivative and thus justified.

Let us note that the non-associative star products induced on D-brane worldvolumes in the presence of non-constant
fields \cite{Cornalba:2001sm} also have several special cyclicity properties, from which one can show them to be
associative up to total derivatives
\cite{Ho:2001qk,Herbst:2001ai,Herbst:2003we}. Our gauge-theoretic star products are essentially the same object, of course
constructed from a very different starting point and acting not on the worldvolume coordinates but the transverse ones. Trying
to understand the D-brane worldvolume origins of our star products, at least at leading order in the deformation,
might help to elucidate the origins of this hidden associativity. 

\subsection{Quasi-Hopf Invariance of the superpotential}

Having expressed the Leigh-Strassler superpotential as a star-product version of that of $\Ncal=4$ SYM, we are
ready to show that it is invariant under our $U_{q,h}[\su(3)]$ quasi-Hopf symmetry. We will show this by
untwisting the $U_{q,h}[\su(3)]$ action to an undeformed $\SU(3)$ action. Since our non-associative setting is
perhaps unfamiliar, we will go through the steps in some detail.

We start by expressing the Leigh-Strassler superpotential (\ref{starW}) as
\be
\Wcal_{LS}=\frac13 \epsilon_{ijk} \Tr[(\Phi^i\astv\Phi^j)\astv \Phi^k]\;,
\ee
where for concreteness we have chosen the left placement of the parentheses, but we might have as well chosen the right one
as they are equal for this cyclically ordered combination.

Let us analyse this star product at the level of the coordinates $z^i$ (of the non-commutative $\Cset^3$ transverse
to the D3-branes in the field theory setup), looking at a single term for simplicity. To make contact with
computations in similar settings (e.g. in \cite{Majid}) we suppress the indices
by introducing the notation $[z_1\otimes z_2]$ for the $9\times 1$ column vector whose components are $z^iz^j$ in the
standard ordering $\{11,12,\ldots\}$ and   $[(z_1\otimes z_2)\otimes z_3]$ for the $27\times 1$ column vector whose components are
$(z^i z^j) z^k$ in the standard ordering $\{111,112,\ldots\}$, and similarly for their star-product versions. So we can
write (\ref{zcubicastL}) as
\be
(z_1\astv z_2)\astv z_3=m\left([(\Delta\otimes \id)(F^{-1})(F^{-1}\otimes 1)] \trr [(z_1\otimes z_2)\otimes z_3]\right)\;.
\ee
Now we would like to transform $z^i\rightarrow z^{i'}=U^{i'}_{\;\;i} z^i$, where $U=e^{iT}$ is an $\SU(3)$ matrix. Assuming (for now)
that we are in a Lie group setting, $U$ will act on the product of two $z$'s as
\be
(z_1z_2)'=m(z_1\otimes z_2)'=m\left([\Delta(U)] \trr [z_1\otimes z_2]\right) =m\left( [\Delta(e^{iT})] \trr [z_1\otimes z_2]\right)\;.
\ee
But, $\Delta$ being the undeformed coproduct, we compute
\be
\Delta(e^{iT})=e^{i\Delta(T)}=e^{i(T\otimes 1+1\otimes T)}=e^{i ~T\otimes 1}e^{i ~1\otimes T}=e^{iT}\otimes e^{iT}=U\otimes U\;,
\ee
so we find
\be
(z_1\otimes z_2)'=m\left([U \otimes U] \trr [z_1\otimes z_2]\right)=m(U\trr z_1 \otimes  U\trr z_2)=(U\trr z_1)\cdot (U\trr z_2)\;.
\ee
Converting this expression to indices, we of course obtain the familiar $\SU(3)$ transformation
\be
z^{i'} z^{j'}= U^{i'}_{\;i}z^i U^{j'}_{\;j}z^j\;.
\ee
 If we now act on three coordinates, we proceed similarly\footnote{The parentheses are of course
  irrelevant in this associative setting but are kept to motivate the twisted discussion to follow.}
\be
((z_1z_2)z_3)'=m((z_1\otimes z_2)\otimes z_3)'=m\left([(\Delta\otimes \id)\Delta(U)] \trr [(z_1\otimes z_2)\otimes z_3]\right)\;.
\ee
We compute
\small
\be
(\Delta\otimes \id)\Delta(U)=(\Delta\otimes \id) e^{i(T\otimes 1+1\otimes T)}=e^{i(\Delta(T)\otimes 1+\Delta(1)\otimes T)}=e^{i(T\otimes 1\otimes1+1\otimes T\otimes 1+1\otimes 1\otimes T)}=U\otimes U\otimes U.
\ee
\normalsize
So in indices we again obtain the usual transformation:
\be
((z^iz^j)z^k)'=(U^{i'}_{\;i}z^iU^{j'}_{\;j}z^j) U^{k'}_{\;k}z^k.
\ee
So far, we have just written down a Lie group action on its representation space in far more detail than necessary. 
However, when we deform the (universal enveloping algebra of our) Lie algebra by twisting, we will need to repeat the
above while taking into account the twisted coproducts $\Delta_F=F\Delta F^{-1}$, as well as the corresponding star products on the
representation space, so the computations are less trivial. Let us start by checking how the quadratic
$\astv$ -product relation transforms under the action of $U$:
\be
\begin{split}
(z_1\astv z_2)'&=m_F([\Delta_F(U)] \trr [z_1\otimes z_2])=m(F^{-1}(F\Delta(U) F^{-1}) \trr [z_1\otimes z_2])\\
&=m(\Delta(U) F^{-1} \trr [z_1\otimes z_2])=(U F^{-1}_{(1)}\trr z_1)\cdot
(U F^{-1}_{(2)}\trr z_2)
\end{split}
  \ee
which, in indices, tells us that
\be
z^{i'}\astv z^{j'}=\sum U^{i'}_{\;i''} [F^{-1}_{(1)}]^{i''}_{\;i} z^i\, U^{j'}_{\;j''} [F^{-1}_{(2)}]^{j''}_{\;j} z^j
=U^{i'}_{\;i''} U^{j'}_{\;j''} [F^{-1}]^{i''\;j''}_{\;i\;\;\;j}z^i z^j=U^{i'}_{\;i} U^{j'}_{\;j} (z^i\astv z^j)\;.
\ee
So the star product, acted on by the twisted coproduct of an $\SU(3)$ element,
transforms in the same way as the undeformed product acted on by the usual Lie algebra coproduct.
We emphasise that the $U$'s appearing in the relation above are just $\SU(3)$ matrices. Only the
way by which they act on multiple copies of the algebra has been twisted. 

We can proceed similarly for three copies of the algebra, where we can write:
\be
U\trr ((z_1\astv z_2)\astv z_3) =U\trr m_F([(z_1\otimes z_2)\otimes z_3])=m_F((\Delta_F\otimes \id) \Delta_F(U)\trr [(z_1\otimes z_2)\otimes z_3])\;.
\ee
But we can compute (adapting a computation in \cite{Majid} to the exponential of an algebra element):
\be
\begin{split}
(\Delta_F\otimes \id)\Delta_F(U)&=(\Delta_F\otimes \id)(F\Delta(e^{iT})F^{-1})=
(\Delta_F\otimes \id)(e^{i f}e^{i(T\otimes 1+1\otimes T)}e^{-i f})\\
&=(F\otimes 1)(\Delta\otimes \id)(e^{i f}e^{i(T\otimes 1+1\otimes T)}e^{-i f})(F^{-1}\otimes 1)\\
&=(F\otimes 1)(e^{i (\Delta\otimes \id) f}e^{i(T\otimes1\otimes 1+1\otimes T\otimes 1+1\otimes 1\otimes T)}e^{-i (\Delta\otimes \id) f})(F^{-1}\otimes 1)\\
&=F_{12}(\Delta\otimes \id)(F) ~(\Delta\otimes \id)\Delta(U) ~ (\Delta\otimes \id)(F^{-1}) F^{-1}_{12}\;.
\end{split}
\ee
We see that the iterated action of the twisted coproduct of $U$ can be expressed as a \emph{twisted} iterated
action of the \emph{untwisted} coproduct of $U$.

Given the above and the definition (\ref{zcubicastL}), it is now easy to show covariance of the group action on the cubic star product:
 \be
 \begin{split}
 ((z_1\astv z_2)\astv z_3)' &=
 m_F([(\Delta_F\otimes \id)\Delta_F(U)] \trr [(z_1\otimes z_2)\otimes z_3])\\
 &=m\left([(\Delta\otimes \id)(F^{-1})(F^{-1}\otimes 1) F_{12}(\Delta\otimes \id)(F) ~(\Delta\otimes \id)\Delta(U)\right.\\
&\qquad\quad \left. \cdot (\Delta\otimes \id) (F^{-1}) F^{-1}_{12}] \trr  [(z_1\otimes z_2)\otimes z_3]\right)\\
 & =m\left([(\Delta\otimes \id)\Delta(U) ~ (\Delta\otimes \id)(F^{-1}) F^{-1}_{12}]\trr  [(z_1\otimes z_2)\otimes z_3]\right)\\
 &   = \sum (U [F_{3,L}]_{(1)}\trr z_1) \cdot (U [F_{3,L}]_{(2)}\trr z_2) \cdot (U [F_{3,L}]_{(3)}\trr z_3)\;,
\end{split}
\ee
which in indices reads
\be
(z^{i'}\astv z^{j'})\astv z^{k'}= U^{i'}_{\;i} U^{j'}_{\;j} U^{k'}_{\;k}((z^i\astv z^j)\astv z^k)\;.
\ee
Again, we find that the cubic star product transforms in the usual way under the undeformed $\SU(3)$.
Reverting to gauge theoretic notation, we can now write
\be \label{invariance}
\begin{split}
   W'_{LS}&=\frac13\epsilon_{i'j'k'}\Tr[(\Phi^i\astv\Phi^j)\astv \Phi^k]'= 
   \frac13 \epsilon_{i'j'k'} U^{i'}_{\;i} U^{j'}_{\;j} U^{k'}_{\;k}\Tr[(\Phi^i\astv\Phi^j)\astv \Phi^k]\\
   &=   \det(U) \epsilon_{ijk} \frac13 \Tr[(\Phi^i\astv\Phi^j)\astv \Phi^k] =W_{LS}
\end{split}
   \ee
which shows the $U_{q,h}[\su(3)]$ invariance of the superpotential. It might appear that we are
claiming $\SU(3)$ invariance, but that is not so, as the true action of $U$ on the fields is via the
twisted coproduct $\Delta_F$ of the $U_{q,h}[\su(3)]$ algebra. What (\ref{invariance}) is stating is
that the action of $\Delta_F$ can be untwisted to a usual $\SU(3)$ action, so that we can make use of
the $\SU(3)$ invariant $\epsilon_{ijk}$ tensor.\footnote{This type of relation between the twisted and
  untwisted algebra (or two different twisted algebras) has had several applications, see e.g. the
  discussions of twist equivalence in \cite{Alekseev:1999bs} as well as \cite{Grosse:2000gd}.}
Certainly when expanding out the star products to
write the superpotential as in (\ref{LSW}), the $\SU(3)$ symmetry is hidden.\footnote{Although one should still
  be able to show  $\SU(3)_{q,h}$ invariance, since instead of $\epsilon_{ijk}$ one writes the superpotential in terms of
  an $E_{ijk}$, which is
an $\SU(3)_{q,h}$-invariant tensor: this is precisely the approach taken in \cite{Mansson:2008xv}.}

\subsection{Antiholomorphic and mixed relations} \label{MixedRelations}

The above discussion revolved around reproducing the superpotential of the $(q,h)$-deformed theories and thus took
place purely in the holomorphic sector spanned by the coordinates $z^i$. In order to make a general
statement about the gauge theory, however, we need to also consider how the twist affects the antiholomorphic
and mixed sectors. An {\it ab initio} approach to this question would involve extending the above structures to
the full $\SO(6)$ scalar sector and defining the star product through the appropriate ($36\times 36$) twist using
(\ref{starprod}). We will, however, take a shortcut by proposing a set of star-product relations which reproduces the
expected mixed quantum plane relations. 

A consistent set of quantum plane relations for the $(q,h)$-deformation was proposed in \cite{Mansson:2008xv}. Adapting
those relations to the current context (where the non-commutativity is governed by the star product), we write them as
\be \label{quantumplaneall}
\begin{split}
R^{i\;j}_{\;k\;l}z^k \star z^l=z^j\star z^i \;,\;\; \zb_k\star \zb_l R^{k\;l}_{\;i\;j}=\zb_j\star \zb_i\;, \\
R^{j\;l}_{\;k\;i} \zb_l \star z^k=z^j\star \zb_i \;,\;\; \widetilde{R}^{i\;l}_{\;k\;j}z^k\star \zb_l=\zb_j \star z^i\;,
\end{split}
\ee
where $\widetilde{R}$ is the \emph{second inverse} of $R$, satisfying
\be \label{Rsecondinv}
\widetilde{R}^{i \;n}_{\;m\;j} R^{m\,k}_{\;\,l\;\;n}=R^{i\;n}_{\;m\;j} \widetilde{R}^{m\,k}_{\;\,l\;\;n}=\delta^i_{\;l}\delta^k_{\;j}\;.
\ee
The second inverse of the $R$-matrix can be constructed from the original one by a procedure of transposing
in the second space, inverting and transposing again (see e.g. \cite{Majid} for a proof). The second inverses
of the other tensors discussed below will also be defined as in (\ref{Rsecondinv}).

Our task is now to find a set of definitions for the mixed star products which reproduces the above relations.
We propose the following extension of (\ref{starindices}):
\be \label{starprodall}
\begin{split}
z^i\star z^j&=F^{i\;j}_{\;k\;l}z^k z^l\;,\;\;\zb_j\star \zb_i=\zb_l\zb_k F^{l\;k}_{\;j\;i} \;,\\
z^i \star \zb_j&=z^k F^{l\;i}_{\;j\;k}\zb_l\;,\;\; \zb_i\star z^j=\zb_l G^{l\;j}_{\;i\;k} z^k\;.
\end{split}
\ee
The holomorphic case was already discussed in section \ref{qhstarsec}, and the anti-holomorphic one follows straightforwardly by
hermitian conjugation. The mixed relations involve a new tensor $G$, whose second inverse $\widetilde{G}$ provides a
second factorisation of the $R$-matrix (analogous to the normal twist-factorisation (\ref{factorisingtwist})) as
\be \label{RtGF}
R^{i\;j}_{\;k\;l}=\widetilde{G}^{j\;m}_{\;n\;k} F^{n\;i}_{\;l\;m}\;.
\ee
The $G$-tensor whose second inverse satisfies this relation is given in appendix \ref{appGtensor}. It turns out that the same $G$-tensor also
factorises the second inverse of the $R$-matrix as
\be
\widetilde{R}^{i\;j}_{\;m\;n}=\widetilde{F}^{j\;l}_{\;k\;m} G^{\;k\;i}_{\;n\;l}\;.
\ee
It can be checked that, taken together, these two expressions imply (\ref{Rsecondinv}). 

Given the above two factorisations of the $R$-matrix, we can now easily see that the definitions (\ref{starprodall}) lead to the quantum plane relations (\ref{quantumplaneall}). For instance,
we can check:
\be
z^j\star\zb_i=z^l\zb_k F^{k\;j}_{\;i\;\;l}=\zb_k z^l F^{k\;j}_{\;i\;\;l}=\zb_m \star z^n\widetilde{G}^{m\;l}_{\;k\;n}F^{k\;j}_{\;i\;l}=\zb_m\star z^n R^{j\,m}_{\;n\;i}\;,
\ee
where we used that
\be
\zb_i\star z^j= G^{l\;j}_{\;i\;k} \zb_l z^k \;\Rightarrow \; \zb_m z^n=\widetilde{G}^{i \;n}_{\;m \;j} \zb_i\star z^j\;.
\ee
Similarly we can show the other mixed quantum plane relation. We conclude that the definitions (\ref{starprodall}) are consistent
with our expectations. As discussed, it would be important to establish them more formally starting from the $36\times 36$ $R$-matrix
of the full $\SO(6)$ scalar sector, which should also provide additional insight into the origin of the $G$-tensor.

\subsection{Twist-invariance of the kinetic terms} \mbox{}

We are now ready to discuss the physical interpretation of our (conjectural) mixed star product relations (\ref{starprodall}).
First, we define the corresponding inverse star products as:
\be
\begin{split}
z^i\astv z^j&=F^{j\;i}_{\;l\;k}z^k z^l\;,\;\;\zb_j\astv \zb_i=\zb_l\zb_k F^{k\;l}_{\;i\;j} \;,\\
z^i \astv \zb_j&=z^k \widetilde{F}^{l\;i}_{\;j\;k}\zb_l\;,\;\; \zb_i\astv z^j=\zb_k \widetilde{G}^{k\,j}_{\;i\;l} z^l\;.
\end{split}
\ee
As in the holomorphic sector, we can use the $\astv$ -product to deform any $\Ncal=4$ SYM expression
involving mixed products of the chiral superfields to the corresponding expression in the Leigh-Strassler
theories. Crucially, even though the individual mixed relations are not particularly transparent (and we
refrain from writing them down), it turns out that the traces (in the quantum algebra) of the mixed star products
evaluate simply to:
\be
\zb_1 \astv z^1+\zb_2\astv z^2+\zb_3 \astv z^3=\zb_1  z^1+\zb_2 z^2+\zb_3 z^3
\ee
as well as
\be
z^1 \astv \zb_1+z^2\astv \zb_2+z^3 \astv \zb_3=z^1 \zb_1+z^2 \zb_2+z^3 \zb_3
\ee
which is of course required by compatibility with the gauge theory trace (when interpreting the $z$'s as chiral superfields). 

This simple result tells us that in the K\"ahler part of the gauge theory action we simply have\footnote{Note that the $e^{gV}$ terms (which couple the scalars to the gauge field), being $\SU(3)$ singlets, will also be singlets in the quasi-Hopf algebra and do not enter this discussion.} 
\be
\overline{\Phi}_i\astv \Phi^i=\overline{\Phi}_i\Phi^i \;.
\ee
So the kinetic part of the action is invariant under the twist.
This confirms that the quasi-Hopf deformation we are considering is a purely superpotential deformation,
as it should be in order to correspond to the Leigh-Strassler marginal deformations.

Of course, when (for instance) relating observables of $\Ncal=4$ SYM to those in the Leigh-Strassler theories
one might have to consider other mixed star product expressions, where the above simplification does not
occur and there will be nontrivial dependence on the $q,h$ parameters.

\section{Conclusions}

In this work we constructed a Drinfeld twist which relates the algebraic structure of $\Ncal=4$ SYM to that
of its $\Ncal=1$ marginal deformations. As anticipated in \cite{Dlamini:2016aaa}, the resulting structure
is that of a quasi-Hopf algebra. So one can think of the Leigh-Strassler deformations as twisting the
$\SU(3)\times\Urm(1)_R$ global Lie-algebraic symmetry of $\Ncal=4$ SYM to a global quasi-Hopf symmetry
$U_{q,h}[\su(3)]\times\Urm(1)_R$. We showed how one can perform explicit computations in this quasi-Hopf setting
by studying the coassociator and demonstrating that the quasi-Hopf analogue of the Yang-Baxter Equation is satisfied.  

Using the twist, we also introduced a star product which relates the Leigh-Strassler lagrangian to that of
$\Ncal=4$ SYM, as well as its inverse, which deforms the $\Ncal=4$ SYM lagrangian to the Leigh-Strassler one.
This generalises previously known star products \cite{Lunin:2005jy,Bundzik2007,Dlamini:2016aaa} which were only applicable to
integrable cases, such as the $\beta$ or $w$-deformation. 

An immediate issue with the above claim of a quantum group global symmetry in QFT has to do with statistics.
One of the important properties of the trivial Lie algebra coproduct is that it admits a natural action of the symmetric
group, so that it makes sense to consider symmetric or antisymmetric wavefunctions for multi-particle states.
This is no longer the case when one deforms the coproduct to obtain a Hopf algebra. Fortunately, it has been
shown in \cite{Fiore:1996tr, Fiore:1996pm}
that for quantum groups derived by twisting Lie algebraic structures (which is precisely our case) it is
possible to adjust the symmetric group action in order to make sense of particle statistics in precisely
the same way as in the undeformed theory. So, interpreted with due care, our quasi-Hopf symmetry is not in
conflict with the principles of Quantum Field Theory.

As we saw in section \ref{cyclicstar}, our star product produces the Leigh-Strassler superpotential with
the appropriate coefficient to satisfy the planar limit of the one-loop conformality condition (\ref{conformalconstraint}).
It would be interesting to study whether non-planar corrections as well as gauge coupling corrections can
be incorporated in the star product so that it can reproduce the full conformal constraint at any given loop
order. 

The uncovering of the quasi-Hopf global symmetry of these $\Ncal=1$ SYM theories can potentially lead to several
useful applications. For instance, one can imagine adapting our star product in order to push the twistor amplitude
computations in \cite{Kulaxizi:2004pa} to higher orders in the deformation parameters. One could similarly
look at higher-loop amplitudes in the gauge theory as was done for real-$\beta$ in \cite{Khoze:2005nd}. One should also
revisit the study of \cite{Mansson:2008xv}, where associativity was imposed by hand, from this new perspective, and understand
whether the algebra defined through the FRT relations (but now allowed to be quasi-associative) can be consistent
at higher levels despite the failure of the YBE.

Clearly, this hidden symmetry of the Leigh-Strassler theories would be expected to lead to conserved currents
and (for instance) relations between observables, which would not be evident if one were to focus only on the discrete
$\Delta_{27}$ symmetry. One could also ask similar questions regarding the spectrum, and for instance revisit studies of
chiral primaries in the Leigh-Strassler theories making use of the discrete symmetries (e.g. \cite{Mauri:2006uw,Madhu:2007ew})
from the quantum algebra perspective. Note that in recent work \cite{Beisert:2017pnr,Beisert:2018zxs,Beisert:2018ijg},  it was
established that a different Hopf algebra, the Yangian, acts as a symmetry on the action of $\Ncal=4$ SYM
as well as that of the $\beta$-deformed theory \cite{Garus:2017bgl}, and its presence indeed led to Slavnov-Taylor relations
between correlation functions. (Of course, given the twist relating our quantum symmetry to the undeformed $\SU(3)$ symmetry
of $\Ncal=4$ SYM, one expects the corresponding relations in our case to be much simpler than those for the Yangian.)

In this work we have only considered gauge-theoretic deformations which preserve $\Ncal=1$ superconformal
invariance. There exists a much larger class of deformations, such as for instance the $\gamma_i$
deformations \cite{Frolov:2005dj} which involve the full $\SO(6)$ symmetry group of the $\Ncal=4$ SYM scalars
instead of just the $\SU(3)$ symmetry of the chiral superfields. Although non-supersymmetric, these more general
deformations admit very interesting limits such as the (non-unitary, but integrable) theories introduced in
\cite{Gurdogan:2015csr}, and it would be relevant to understand these deformations from a quantum group perspective.

An intriguing possibility is that, if one is able to identify how the quasi-Hopf symmetry acts on the dual gravity
side, one might use the quasi-Hopf twist as a solution-generating technique in order to construct the supergravity
dual. For Hopf twists, this has been shown to work in \cite{Dlamini:2016aaa}, where the supergravity dual of
the $w$-deformation was constructed without going through the route of TST transformations. Rather, the star
product was applied to the pure spinors of the generalised geometry description of flat space and was shown
to lead to the so-called NS-NS precursor of the dual geometry  \cite{Lunin:2005jy}, i.e. the background which leads to
the actual dual geometry on inserting D-branes at the origin and taking the near-horizon limit. Of course,
there are several challenges ahead in extending this approach to general $(q,h)$, most notably the issue of
non-associativity. 

We should emphasise that the dual IIB background is a smooth geometry, and not noncommutative in any way. The
quasi-Hopf algebra will only appear in the open-string metric, which is of course the one seen by the scalars of
the gauge theory. As is standard, the non-commutativity (and here also non-associativity) seen by the open
strings should be exchanged with the B-field and RR fields seen by the closed strings via the Seiberg-Witten open/closed
mapping \cite{Seiberg:1999vs}, appropriately generalised to non-constant fields. Indeed, perhaps a useful (but
at this stage very imprecise) way to think of our star product is as the all-orders generalisation of the
non-associative star products appearing on the world-volumes of D-branes in the presence of non-constant background fields
\cite{Cornalba:2001sm,Ho:2001qk,Herbst:2001ai,Herbst:2003we}. We should note that the open/closed mapping has already
been used to construct the dual background
to the Leigh-Strassler theories \cite{Kulaxizi:2006pp,Kulaxizi:2006zc}, although (apart from the real-$\beta$ case)
the construction only worked to second order in the deformation parameter because of ambiguities related to
non-associativity. One could hope that the improved understanding of non-associativity achieved here
might help to extend that construction to higher orders.

 Recently, a large class of integrable deformations of the $\AdS_5\times \Srm^5$ sigma-model
have been obtained through Yang-Baxter sigma model deformations \cite{Klimcik:2002zj,Klimcik:2008eq}, based
on solutions of a modified \cite{Delduc:2013qra} or unmodified \cite{Kawaguchi:2014qwa} classical Yang-Baxter equation.
The latter have been interpreted in terms of Drinfeld twists in \cite{vanTongeren:2015uha,vanTongeren:2016eeb}
(see also \cite{Araujo:2017jap}
for further discussion of Drinfeld twists in this context). It would be interesting to understand how our twists
(or at least their classical counterparts) might fit into the Yang-Baxter sigma model framework.

In studying the twist and its implications, we have been guided by the gauge theory, and in particular the
requirement of exact four-dimensional superconformal invariance, without any special regard to
integrability. But, of course, one of the main questions motivating this work has been to understand how the 
integrability properties of planar $\Ncal=4$ SYM change as one deforms away from maximal supersymmetry.
In particular, what makes the $\beta$-deformation so special compared to the more general $(q,h)$-deformations?
At some level, this has already been answered by studying which special cases of the $(q,h)$ $R$-matrix satisfy
the YBE \cite{Bundzik:2005zg}. However, here we have found that \emph{all} $(q,h)$ $R$-matrices satisfy a modified form of the YBE. So
quasitriangularity, a defining feature of integrability, is still present for the full Leigh-Strassler
theory, though within the wider context of quasi-Hopf algebras. Of course, our discussion has been
in the quantum group limit of infinite spectral parameter. But it turns out that the spectral parameter
can be reintroduced with very little effort: Given the factorisation $R_{q,h}= F_{21} (I\otimes I) F^{-1}$, 
we can define a spectral-parameter-dependent $R_{q,h}(u)$-matrix starting from the XXX Heisenberg $R(u)$-matrix as:
\be
R_{q,h}(u)=F_{21}\left(u I\otimes I\!+\!iP\right)F^{-1}=u R_{q,h}+i(P F_{12} P)PF^{-1}=u R_{q,h}+iP F_{12}F^{-1}
=uR_{q,h}+iP.
\ee
So the twist only affects the identity part of the $R$-matrix, just as in \cite{Beisert:2005if} for the $\gamma_i$ twists.
It follows (and can be explicitly shown) that the qHYBE will hold with spectral parameter, with the same coassociator:
\be
R_{12}(u)\Phi_{312}R_{13}(u+v)\Phi^{-1}_{132} R_{23}(v)\Phi_{123}=\Phi_{321}R_{23}(v)\Phi^{-1}_{231}R_{13}(u+v)\Phi_{213}R_{12}(u)\;.
\ee 
Clearly, the study of how the algebraic Bethe ansatz construction would apply in our current quasi-Hopf setting,
and whether that might lead to a re-evaluation of the integrability properties of the Leigh-Strassler theories,
is a very interesting open question.

\paragraph{Acknowledgments} We would like to thank Robert de Mello Koch and Manuela Kulaxizi for very useful
discussions. KZ would also like to thank the organisers and participants of the 10$^{\text{th}}$ Joburg Meeting on String
Theory, the 5$^\text{th}$ Athens Xmas theoretical physics workshop and the 3$^\text{rd}$ Mandelstam Theoretical Physics School
and Workshop, where this work was presented, for useful input and feedback. The research of KZ was supported
by the National Research Foundation of South Africa through grants CSUR-93735 and Incentive-103895. HD was
supported through a PhD bursary by the National Institute for Theoretical Physics.

\appendix
   
   \section{The imaginary-$\beta$ coassociator} \label{imbcoassociator}

   In this appendix we list the non-zero components of the imaginary-$\beta$ (real $q$) coassociator. We show
   only the components with the first index being $1$, as the remaining components can be obtained by
   cyclically shifting the indices. 

   \tiny

\be
\Phi^{1\;1\;1}_{\;1\;1\;1}=1
\ee
\be
\Phi^{1\;1\;2}_{\;1\;1\;2}=\Phi^{1\;2\;2}_{\;1\;2\;2}=\frac{q \left(\sqrt{2} \sin \left(2 \sqrt{2}\,\alpha\right)+4\right)
  -2 \sqrt{2} \sin \left(\sqrt{2}\,\alpha\right)+4 \cos \left(\sqrt{2}\,\alpha\right)}{4 \sqrt{2} \sqrt{q^2+1}}
\ee
\be
\Phi^{1\;1\;3}_{\;1\;1\;3}=\Phi^{1\;3\;3}_{\;1\;3\;3}=\frac{2 \sqrt{2} \,q \sin \left(\sqrt{2}\,\alpha\right)+4\,q \cos \left(\sqrt{2}\,\alpha\right)
  -\sqrt{2} \sin \left(2 \sqrt{2}\,\alpha\right)+4}{4 \sqrt{2} \sqrt{q^2+1}}
\ee
\be
\Phi^{1\;1\;2}_{\;1\;2\;1}=\Phi^{1\;3\;1}_{\;3\;1\;1}=\frac{q^2 \cos \left(2 \sqrt{2}\,\alpha\right)-4 \sqrt{2}\,q \sin \left(\sqrt{2}\,\alpha\right)+q^2-2}{4 \left(q^2+1\right)}
\ee
\be
\Phi^{1\;1\;3}_{\;1\;3\;1}=\Phi^{1\;2\;1}_{\;2\;1\;1}=\frac{4 \sqrt{2}\,q \sin \left(\sqrt{2}\,\alpha\right)+\cos \left(2 \sqrt{2}\,\alpha\right)-2\,q^2+1}{4 \left(q^2+1\right)}
\ee
\be
\Phi^{1\;2\;1}_{\;1\;1\;2}=\Phi^{1\;2\;2}_{\;2\;1\;2}=\frac{2 \sqrt{2}\,q \sin \left(\sqrt{2}\,\alpha\right)-4\,q \cos \left(\sqrt{2}\,\alpha\right)+\sqrt{2} \sin \left(2 \sqrt{2}\,\alpha\right)+4}{4 \sqrt{2} \sqrt{q^2+1}}
\ee
\be
\Phi^{1\;2\;1}_{\;1\;2\;1}=\Phi^{1\;3\;1}_{\;1\;3\;1}=\frac{4\,q^2 \sin \left(\sqrt{2}\,\alpha\right)+2 \sqrt{2} \left(q^2+1\right) \cos \left(\sqrt{2}\,\alpha\right)+\sqrt{2}\,q \cos \left(2 \sqrt{2}\,\alpha\right)-4 \sin \left(\sqrt{2}\,\alpha\right)+3 \sqrt{2}\,q}{4 \sqrt{2} \left(q^2+1\right)}
\ee
\be
\Phi^{1\;2\;3}_{\;1\;2\;3}=\Phi^{1\;3\;2}_{\;1\;3\;2}=\frac{\sqrt{3} \left(q^2-1\right) \sin \left(2 \sqrt{3}\,\alpha\right)+\left(q^2+1\right) \cos \left(2 \sqrt{3}\,\alpha\right)+2 \left(q^2+3\,q+1\right)}{6 \left(q^2+1\right)}
\ee
\be
\Phi^{1\;2\;3}_{\;1\;3\;2}=\Phi^{1\;3\;2}_{\;3\;1\;2}=\frac{-2 \sqrt{3}\,q \sin \left(2 \sqrt{3}\,\alpha\right)+3\,q^2-3}{6 \left(q^2+1\right)}
\ee
\be
\Phi^{1\;3\;1}_{\;1\;1\;3}=\Phi^{1\;3\;3}_{\;3\;1\;3}= -\frac{q \left(\sqrt{2} \sin \left(2 \sqrt{2}\,\alpha\right)-4\right)+2 \sqrt{2} \sin \left(\sqrt{2}\,\alpha\right)+4 \cos \left(\sqrt{2}\,\alpha\right)}{4 \sqrt{2} \sqrt{q^2+1}}
  \ee
  \be
  \Phi^{1\;3\;2}_{\;1\;2\;3}= \Phi^{1\;2\;3}_{\;2\;1\;3}=\frac{2 \sqrt{3}\,q \sin \left(2 \sqrt{3}\,\alpha\right)-3\,q^2+3}{6 \left(q^2+1\right)}
  \ee
\be
\Phi^{1\;1\;2}_{\;2\;1\;1}=\Phi^{1\;1\;3}_{\;3\;1\;1}=\frac{4\,q^2 \sin \left(\sqrt{2}\,\alpha\right)-2 \sqrt{2} \left(q^2+1\right) \cos \left(\sqrt{2}\,\alpha\right)+\sqrt{2}\,q \cos \left(2 \sqrt{2}\,\alpha\right)-4 \sin \left(\sqrt{2}\,\alpha\right)+3 \sqrt{2}\,q}{4 \sqrt{2} \left(q^2+1\right)}
\ee
      \be
  \Phi^{1\;2\;2}_{\;2\;2\;1}=\Phi^{1\;3\;3}_{\;3\;3\;1}=-{{\sin ^2\left(\sqrt{2}\,\alpha\right)}\over{2}}
  \ee
       \be
  \Phi^{1\;3\;2}_{\;3\;2\;1}=\Phi^{1\;2\;3}_{\;2\;3\;1}={{2\,\sin ^2\left(\sqrt{3}\,\alpha\right)}\over{3}}
  \ee
\be
  \Phi^{1\;3\;2}_{\;2\;1\;3}=\Phi^{1\;2\;3}_{\;3\;1\;2}=\frac{-\sqrt{3} \left(q^2-1\right) \sin \left(2 \sqrt{3}\,\alpha\right)+\left(q^2+1\right) \cos \left(2 \sqrt{3}\,\alpha\right)+2 \left(q^2-3\,q+1\right)}{6 \left(q^2+1\right)}
   \ee
   \normalsize
   
Here $\alpha=\arccos((1+q)/(\sqrt{2}\sqrt{1+q^2}))$ as defined in (\ref{alphaq})

\normalsize

\section{The left and right cubic matrices} \label{matrices}
In this appendix we record the matrices $i(\Delta\otimes \id)(f)$ and $i(\id\otimes \Delta)(f)$ which need to be
exponentiated to obtain $(\Delta\otimes\id)(F)$, $(\id\otimes\Delta)(F)$ and their inverses. We have redefined $\alpha_{h_r}$
and $\alpha_{h_i}$ as $\alpha_{h_r}= \half(\alpha_h^++\alpha_h^-)$ and $\alpha_{h_i}=\frac{1}{2i}(\alpha_h^+-\alpha_h^-)$. 
         {\tiny
  \be\nonumber
\begin{split}
  \quad  &i(\Delta\otimes \id)f=\\
&    \left(
   \arraycolsep=0.35pt
\begin{array}{ccccccccccccccccccccccccccc}
 0 & 0 & 0 & 0 & 0 & {\alpha_h^-} & 0 & -{\alpha_h^-} & 0 & 0 & 0 & {\alpha_h^-} & 0 & 0 & 0 & 0 & 0 & 0 & 0 & -{\alpha_h^-} & 0 & 0 & 0 & 0 & 0 & 0 & 0 \\
 0 & 2 i {\alpha_{\beta}^r} & 0 & {\alpha_\beta^i} & 0 & 0 & 0 & 0 & -{\alpha_{h}^+} & {\alpha_\beta^i} & 0 & 0 & 0 & 0 & 0 & 0 & 0 & 0 & 0 & 0 & -{\alpha_{h}^+} & 0 & 0 & 0 & 0 & 0 & 0 \\
 0 & 0 & -2 i {\alpha_{\beta}^r} & 0 & {\alpha_{h}^+} & 0 & -{\alpha_\beta^i} & 0 & 0 & 0 & {\alpha_{h}^+} & 0 & 0 & 0 & 0 & 0 & 0 & 0 & -{\alpha_\beta^i} & 0 & 0 & 0 & 0 & 0 & 0 & 0 & 0 \\
 0 & -{\alpha_\beta^i} & 0 & -i {\alpha_{\beta}^r} & 0 & 0 & 0 & 0 & {\alpha_{h}^+} & 0 & 0 & 0 & 0 & 0 & {\alpha_h^-} & 0 & 0 & 0 & 0 & 0 & 0 & 0 & -{\alpha_h^-} & 0 & 0 & 0 & 0 \\
 0 & 0 & -{\alpha_h^-} & 0 & i {\alpha_{\beta}^r} & 0 & {\alpha_h^-} & 0 & 0 & 0 & 0 & 0 & {\alpha_\beta^i} & 0 & 0 & 0 & 0 & 0 & 0 & 0 & 0 & 0 & 0 & -{\alpha_{h}^+} & 0 & 0 & 0 \\
 -{\alpha_{h}^+} & 0 & 0 & 0 & 0 & 0 & 0 & {\alpha_\beta^i} & 0 & 0 & 0 & 0 & 0 & {\alpha_{h}^+} & 0 & 0 & 0 & 0 & 0 & 0 & 0 & -{\alpha_\beta^i} & 0 & 0 & 0 & 0 & 0 \\
 0 & 0 & {\alpha_\beta^i} & 0 & -{\alpha_{h}^+} & 0 & i {\alpha_{\beta}^r} & 0 & 0 & 0 & 0 & 0 & 0 & 0 & 0 & 0 & 0 & {\alpha_h^-} & 0 & 0 & 0 & 0 & 0 & 0 & 0 & -{\alpha_h^-} & 0 \\
 {\alpha_{h}^+} & 0 & 0 & 0 & 0 & -{\alpha_\beta^i} & 0 & 0 & 0 & 0 & 0 & 0 & 0 & 0 & 0 & {\alpha_\beta^i} & 0 & 0 & 0 & 0 & 0 & 0 & 0 & 0 & 0 & 0 & -{\alpha_{h}^+} \\
 0 & {\alpha_h^-} & 0 & -{\alpha_h^-} & 0 & 0 & 0 & 0 & -i {\alpha_{\beta}^r} & 0 & 0 & 0 & 0 & 0 & 0 & 0 & {\alpha_{h}^+} & 0 & 0 & 0 & 0 & 0 & 0 & 0 & -{\alpha_\beta^i} & 0 & 0 \\
 0 & -{\alpha_\beta^i} & 0 & 0 & 0 & 0 & 0 & 0 & 0 & -i {\alpha_{\beta}^r} & 0 & 0 & 0 & 0 & {\alpha_h^-} & 0 & -{\alpha_h^-} & 0 & 0 & 0 & {\alpha_{h}^+} & 0 & 0 & 0 & 0 & 0 & 0 \\
 0 & 0 & -{\alpha_h^-} & 0 & 0 & 0 & 0 & 0 & 0 & 0 & i {\alpha_{\beta}^r} & 0 & {\alpha_\beta^i} & 0 & 0 & 0 & 0 & -{\alpha_{h}^+} & {\alpha_h^-} & 0 & 0 & 0 & 0 & 0 & 0 & 0 & 0 \\
 -{\alpha_{h}^+} & 0 & 0 & 0 & 0 & 0 & 0 & 0 & 0 & 0 & 0 & 0 & 0 & {\alpha_{h}^+} & 0 & -{\alpha_\beta^i} & 0 & 0 & 0 & {\alpha_\beta^i} & 0 & 0 & 0 & 0 & 0 & 0 & 0 \\
 0 & 0 & 0 & 0 & -{\alpha_\beta^i} & 0 & 0 & 0 & 0 & 0 & -{\alpha_\beta^i} & 0 & -2 i {\alpha_{\beta}^r} & 0 & 0 & 0 & 0 & {\alpha_{h}^+} & 0 & 0 & 0 & 0 & 0 & {\alpha_{h}^+} & 0 & 0 & 0 \\
 0 & 0 & 0 & 0 & 0 & -{\alpha_h^-} & 0 & 0 & 0 & 0 & 0 & -{\alpha_h^-} & 0 & 0 & 0 & {\alpha_h^-} & 0 & 0 & 0 & 0 & 0 & {\alpha_h^-} & 0 & 0 & 0 & 0 & 0 \\
 0 & 0 & 0 & -{\alpha_{h}^+} & 0 & 0 & 0 & 0 & 0 & -{\alpha_{h}^+} & 0 & 0 & 0 & 0 & 2 i {\alpha_{\beta}^r} & 0 & {\alpha_\beta^i} & 0 & 0 & 0 & 0 & 0 & {\alpha_\beta^i} & 0 & 0 & 0 & 0 \\
 0 & 0 & 0 & 0 & 0 & 0 & 0 & -{\alpha_\beta^i} & 0 & 0 & 0 & {\alpha_\beta^i} & 0 & -{\alpha_{h}^+} & 0 & 0 & 0 & 0 & 0 & 0 & 0 & 0 & 0 & 0 & 0 & 0 & {\alpha_{h}^+} \\
 0 & 0 & 0 & 0 & 0 & 0 & 0 & 0 & -{\alpha_h^-} & {\alpha_{h}^+} & 0 & 0 & 0 & 0 & -{\alpha_\beta^i} & 0 & -i {\alpha_{\beta}^r} & 0 & 0 & 0 & 0 & 0 & 0 & 0 & {\alpha_h^-} & 0 & 0 \\
 0 & 0 & 0 & 0 & 0 & 0 & -{\alpha_{h}^+} & 0 & 0 & 0 & {\alpha_h^-} & 0 & -{\alpha_h^-} & 0 & 0 & 0 & 0 & i {\alpha_{\beta}^r} & 0 & 0 & 0 & 0 & 0 & 0 & 0 & {\alpha_\beta^i} & 0 \\
 0 & 0 & {\alpha_\beta^i} & 0 & 0 & 0 & 0 & 0 & 0 & 0 & -{\alpha_{h}^+} & 0 & 0 & 0 & 0 & 0 & 0 & 0 & i {\alpha_{\beta}^r} & 0 & 0 & 0 & 0 & {\alpha_h^-} & 0 & -{\alpha_h^-} & 0 \\
 {\alpha_{h}^+} & 0 & 0 & 0 & 0 & 0 & 0 & 0 & 0 & 0 & 0 & -{\alpha_\beta^i} & 0 & 0 & 0 & 0 & 0 & 0 & 0 & 0 & 0 & {\alpha_\beta^i} & 0 & 0 & 0 & 0 & -{\alpha_{h}^+} \\
 0 & {\alpha_h^-} & 0 & 0 & 0 & 0 & 0 & 0 & 0 & -{\alpha_h^-} & 0 & 0 & 0 & 0 & 0 & 0 & 0 & 0 & 0 & 0 & -i {\alpha_{\beta}^r} & 0 & {\alpha_{h}^+} & 0 & -{\alpha_\beta^i} & 0 & 0 \\
 0 & 0 & 0 & 0 & 0 & {\alpha_\beta^i} & 0 & 0 & 0 & 0 & 0 & 0 & 0 & -{\alpha_{h}^+} & 0 & 0 & 0 & 0 & 0 & -{\alpha_\beta^i} & 0 & 0 & 0 & 0 & 0 & 0 & {\alpha_{h}^+} \\
 0 & 0 & 0 & {\alpha_{h}^+} & 0 & 0 & 0 & 0 & 0 & 0 & 0 & 0 & 0 & 0 & -{\alpha_\beta^i} & 0 & 0 & 0 & 0 & 0 & -{\alpha_h^-} & 0 & -i {\alpha_{\beta}^r} & 0 & {\alpha_h^-} & 0 & 0 \\
 0 & 0 & 0 & 0 & {\alpha_h^-} & 0 & 0 & 0 & 0 & 0 & 0 & 0 & -{\alpha_h^-} & 0 & 0 & 0 & 0 & 0 & -{\alpha_{h}^+} & 0 & 0 & 0 & 0 & i {\alpha_{\beta}^r} & 0 & {\alpha_\beta^i} & 0 \\
 0 & 0 & 0 & 0 & 0 & 0 & 0 & 0 & {\alpha_\beta^i} & 0 & 0 & 0 & 0 & 0 & 0 & 0 & -{\alpha_{h}^+} & 0 & 0 & 0 & {\alpha_\beta^i} & 0 & -{\alpha_{h}^+} & 0 & 2 i {\alpha_{\beta}^r} & 0 & 0 \\
 0 & 0 & 0 & 0 & 0 & 0 & {\alpha_{h}^+} & 0 & 0 & 0 & 0 & 0 & 0 & 0 & 0 & 0 & 0 & -{\alpha_\beta^i} & {\alpha_{h}^+} & 0 & 0 & 0 & 0 & -{\alpha_\beta^i} & 0 & -2 i {\alpha_{\beta}^r} & 0 \\
 0 & 0 & 0 & 0 & 0 & 0 & 0 & {\alpha_h^-} & 0 & 0 & 0 & 0 & 0 & 0 & 0 & -{\alpha_h^-} & 0 & 0 & 0 & {\alpha_h^-} & 0 & -{\alpha_h^-} & 0 & 0 & 0 & 0 & 0 \\
\end{array}
\right)
\end{split}
\ee
}  
         {\tiny
  \be\nonumber
\begin{split}
  \quad  & i(\id\otimes\Delta)(f)=\\
&  \left(
   \arraycolsep=0.35pt
  \begin{array}{ccccccccccccccccccccccccccc}
 0 & 0 & 0 & 0 & 0 & 0 & 0 & 0 & 0 & 0 & 0 & {\alpha_h^-} & 0 & 0 & 0 & {\alpha_h^-} & 0 & 0 & 0 & -{\alpha_h^-} & 0 & -{\alpha_h^-} & 0 & 0 & 0 & 0 & 0 \\
 0 & i {\alpha_\beta^r} & 0 & 0 & 0 & 0 & 0 & 0 & 0 & {\alpha_\beta^i} & 0 & 0 & 0 & 0 & 0 & 0 & {\alpha_h^-} & 0 & 0 & 0 & -{\alpha_h^+} & 0 & -{\alpha_h^-} & 0 & 0 & 0 & 0 \\
 0 & 0 & -i {\alpha_\beta^r} & 0 & 0 & 0 & 0 & 0 & 0 & 0 & {\alpha_h^+} & 0 & 0 & 0 & 0 & 0 & 0 & {\alpha_h^-} & -{\alpha_\beta^i} & 0 & 0 & 0 & 0 & -{\alpha_h^-} & 0 & 0 & 0 \\
 0 & 0 & 0 & i {\alpha_\beta^r} & 0 & 0 & 0 & 0 & 0 & {\alpha_\beta^i} & 0 & 0 & 0 & 0 & {\alpha_h^-} & 0 & 0 & 0 & 0 & 0 & 0 & 0 & -{\alpha_h^-} & 0 & -{\alpha_h^+} & 0 & 0 \\
 0 & 0 & 0 & 0 & 2 i {\alpha_\beta^r} & 0 & 0 & 0 & 0 & 0 & {\alpha_\beta^i} & 0 & {\alpha_\beta^i} & 0 & 0 & 0 & 0 & 0 & 0 & 0 & 0 & 0 & 0 & -{\alpha_h^+} & 0 & -{\alpha_h^+} & 0 \\
 0 & 0 & 0 & 0 & 0 & 0 & 0 & 0 & 0 & 0 & 0 & {\alpha_\beta^i} & 0 & {\alpha_h^+} & 0 & 0 & 0 & 0 & 0 & 0 & 0 & -{\alpha_\beta^i} & 0 & 0 & 0 & 0 & -{\alpha_h^+} \\
 0 & 0 & 0 & 0 & 0 & 0 & -i {\alpha_\beta^r} & 0 & 0 & 0 & 0 & 0 & {\alpha_h^+} & 0 & 0 & 0 & 0 & {\alpha_h^-} & -{\alpha_\beta^i} & 0 & 0 & 0 & 0 & 0 & 0 & -{\alpha_h^-} & 0 \\
 0 & 0 & 0 & 0 & 0 & 0 & 0 & 0 & 0 & 0 & 0 & 0 & 0 & {\alpha_h^+} & 0 & {\alpha_\beta^i} & 0 & 0 & 0 & -{\alpha_\beta^i} & 0 & 0 & 0 & 0 & 0 & 0 & -{\alpha_h^+} \\
 0 & 0 & 0 & 0 & 0 & 0 & 0 & 0 & -2 i {\alpha_\beta^r} & 0 & 0 & 0 & 0 & 0 & {\alpha_h^+} & 0 & {\alpha_h^+} & 0 & 0 & 0 & -{\alpha_\beta^i} & 0 & 0 & 0 & -{\alpha_\beta^i} & 0 & 0 \\
 0 & -{\alpha_\beta^i} & 0 & -{\alpha_\beta^i} & 0 & 0 & 0 & 0 & 0 & -2 i {\alpha_\beta^r} & 0 & 0 & 0 & 0 & 0 & 0 & 0 & 0 & 0 & 0 & {\alpha_h^+} & 0 & 0 & 0 & {\alpha_h^+} & 0 & 0 \\
 0 & 0 & -{\alpha_h^-} & 0 & -{\alpha_\beta^i} & 0 & 0 & 0 & 0 & 0 & -i {\alpha_\beta^r} & 0 & 0 & 0 & 0 & 0 & 0 & 0 & {\alpha_h^-} & 0 & 0 & 0 & 0 & 0 & 0 & {\alpha_h^+} & 0 \\
 -{\alpha_h^+} & 0 & 0 & 0 & 0 & -{\alpha_\beta^i} & 0 & 0 & 0 & 0 & 0 & 0 & 0 & 0 & 0 & 0 & 0 & 0 & 0 & {\alpha_\beta^i} & 0 & 0 & 0 & 0 & 0 & 0 & {\alpha_h^+} \\
 0 & 0 & 0 & 0 & -{\alpha_\beta^i} & 0 & -{\alpha_h^-} & 0 & 0 & 0 & 0 & 0 & -i {\alpha_\beta^r} & 0 & 0 & 0 & 0 & 0 & {\alpha_h^-} & 0 & 0 & 0 & 0 & {\alpha_h^+} & 0 & 0 & 0 \\
 0 & 0 & 0 & 0 & 0 & -{\alpha_h^-} & 0 & -{\alpha_h^-} & 0 & 0 & 0 & 0 & 0 & 0 & 0 & 0 & 0 & 0 & 0 & {\alpha_h^-} & 0 & {\alpha_h^-} & 0 & 0 & 0 & 0 & 0 \\
 0 & 0 & 0 & -{\alpha_h^+} & 0 & 0 & 0 & 0 & -{\alpha_h^-} & 0 & 0 & 0 & 0 & 0 & i {\alpha_\beta^r} & 0 & 0 & 0 & 0 & 0 & {\alpha_h^-} & 0 & {\alpha_\beta^i} & 0 & 0 & 0 & 0 \\
 -{\alpha_h^+} & 0 & 0 & 0 & 0 & 0 & 0 & -{\alpha_\beta^i} & 0 & 0 & 0 & 0 & 0 & 0 & 0 & 0 & 0 & 0 & 0 & 0 & 0 & {\alpha_\beta^i} & 0 & 0 & 0 & 0 & {\alpha_h^+} \\
 0 & -{\alpha_h^+} & 0 & 0 & 0 & 0 & 0 & 0 & -{\alpha_h^-} & 0 & 0 & 0 & 0 & 0 & 0 & 0 & i {\alpha_\beta^r} & 0 & 0 & 0 & 0 & 0 & {\alpha_\beta^i} & 0 & {\alpha_h^-} & 0 & 0 \\
 0 & 0 & -{\alpha_h^+} & 0 & 0 & 0 & -{\alpha_h^+} & 0 & 0 & 0 & 0 & 0 & 0 & 0 & 0 & 0 & 0 & 2 i {\alpha_\beta^r} & 0 & 0 & 0 & 0 & 0 & {\alpha_\beta^i} & 0 & {\alpha_\beta^i} & 0 \\
 0 & 0 & {\alpha_\beta^i} & 0 & 0 & 0 & {\alpha_\beta^i} & 0 & 0 & 0 & -{\alpha_h^+} & 0 & -{\alpha_h^+} & 0 & 0 & 0 & 0 & 0 & 2 i {\alpha_\beta^r} & 0 & 0 & 0 & 0 & 0 & 0 & 0 & 0 \\
 {\alpha_h^+} & 0 & 0 & 0 & 0 & 0 & 0 & {\alpha_\beta^i} & 0 & 0 & 0 & -{\alpha_\beta^i} & 0 & -{\alpha_h^+} & 0 & 0 & 0 & 0 & 0 & 0 & 0 & 0 & 0 & 0 & 0 & 0 & 0 \\
 0 & {\alpha_h^-} & 0 & 0 & 0 & 0 & 0 & 0 & {\alpha_\beta^i} & -{\alpha_h^-} & 0 & 0 & 0 & 0 & -{\alpha_h^+} & 0 & 0 & 0 & 0 & 0 & i {\alpha_\beta^r} & 0 & 0 & 0 & 0 & 0 & 0 \\
 {\alpha_h^+} & 0 & 0 & 0 & 0 & {\alpha_\beta^i} & 0 & 0 & 0 & 0 & 0 & 0 & 0 & -{\alpha_h^+} & 0 & -{\alpha_\beta^i} & 0 & 0 & 0 & 0 & 0 & 0 & 0 & 0 & 0 & 0 & 0 \\
 0 & {\alpha_h^+} & 0 & {\alpha_h^+} & 0 & 0 & 0 & 0 & 0 & 0 & 0 & 0 & 0 & 0 & -{\alpha_\beta^i} & 0 & -{\alpha_\beta^i} & 0 & 0 & 0 & 0 & 0 & -2 i {\alpha_\beta^r} & 0 & 0 & 0 & 0 \\
 0 & 0 & {\alpha_h^+} & 0 & {\alpha_h^-} & 0 & 0 & 0 & 0 & 0 & 0 & 0 & -{\alpha_h^-} & 0 & 0 & 0 & 0 & -{\alpha_\beta^i} & 0 & 0 & 0 & 0 & 0 & -i {\alpha_\beta^r} & 0 & 0 & 0 \\
 0 & 0 & 0 & {\alpha_h^-} & 0 & 0 & 0 & 0 & {\alpha_\beta^i} & -{\alpha_h^-} & 0 & 0 & 0 & 0 & 0 & 0 & -{\alpha_h^+} & 0 & 0 & 0 & 0 & 0 & 0 & 0 & i {\alpha_\beta^r} & 0 & 0 \\
 0 & 0 & 0 & 0 & {\alpha_h^-} & 0 & {\alpha_h^+} & 0 & 0 & 0 & -{\alpha_h^-} & 0 & 0 & 0 & 0 & 0 & 0 & -{\alpha_\beta^i} & 0 & 0 & 0 & 0 & 0 & 0 & 0 & -i {\alpha_\beta^r} & 0 \\
 0 & 0 & 0 & 0 & 0 & {\alpha_h^-} & 0 & {\alpha_h^-} & 0 & 0 & 0 & -{\alpha_h^-} & 0 & 0 & 0 & -{\alpha_h^-} & 0 & 0 & 0 & 0 & 0 & 0 & 0 & 0 & 0 & 0 & 0 \\
\end{array}
\right)
\end{split}
\ee
}
         Direct exponentiation of these matrices in Mathematica has proved challenging, so, as explained in section \ref{generalcoassocsec}, we have first
         constructed their eigenvectors in order to use the relation $\exp(M)=V^\dag \exp(D) V$, with $D$ being the diagonalised version of the matrices. The details can be found in the Mathematica worksheet made available with this preprint.

\section{The imaginary-$\beta$ cubic product tensors} \label{imbcubic}

In this appendix we record the explicit expressions of the
combinations of the twists that appear in the cubic star products. 

As in  appendix \ref{imbcoassociator}, we fix the $\Zset_3$
symmetry by choosing the first index of $[F_{3,L}]$ and the last index of $[F_{3,R}]$ to be 1.
The two other equal components are obtained by shifting all indices
up or down. Otherwise, these are the only nonzero
components.  The parameter $\alpha$ is as in (\ref{alphaq}).

\small

\be
   [F_{3,L}]^{1\;1\;1}_{\;1\;1\;1}=[F_{3,R}]^{1\;1\;1}_{\;1\;1\;1}=1
   \ee

\be
   [F_{3,L}]^{1\;1\;2}_{\;1\;1\;2}=[F_{3,L}]^{1\;1\;3}_{\;1\;1\;3}=[F_{3,R}]^{3\;1\;1}_{\;3\;1\;1}=[F_{3,R}]^{2\;1\;1}_{\;2\;1\;1}=\cos \left(\sqrt{2}\,\alpha\right)
   \ee

   \be
      [F_{3,L}]^{1\;1\;2}_{\;1\;2\;1}=-[F_{3,L}]^{1\;1\;3}_{\;3\;1\;1}=[F_{3,R}]^{3\;1\;1}_{\;1\;3\;1}=-[F_{3,R}]^{1\;3\;3}_{\;3\;3\;1}=
-{{q\,\sin \left(\sqrt{2}\,\alpha\right)}\over{\sqrt{q^2+1}}}
   \ee

\be
   [F_{3,L}]^{1\;1\;3}_{\;1\;3\;1}=-[F_{3,L}]^{1\;1\;2}_{\;2\;1\;1}=-[F_{3,R}]^{1\;2\;2}_{\;2\;2\;1}=[F_{3,R}]^{2\;1\;1}_{\;1\;2\;1}={{\sin \left(\sqrt{2}\,
       \alpha\right)}\over{\sqrt{q^2+1}}}
   \ee

   \be
   \begin{split}
     [F_{3,L}]^{1\;2\;1}_{\;1\;1\;2}&=-[F_{3,L}]^{1\;2\;2}_{\;2\;2\;1}==-[F_{3,L}]^{1\;3\;1}_{\;1\;1\;3}=[F_{3,L}]^{1\;3\;3}_{\;3\;3\;1}\\
     &=-[F_{3,R}]^{1\;1\;2}_{\;2\;1\;1}=-[F_{3,R}]^{1\;2\;1}_{\;2\;1\;1}=[F_{3,R}]^{1\;1\;3}_{\;3\;1\;1}=[F_{3,R}]^{1\;3\;1}_{\;3\;1\;1}={{\sin \left(\sqrt{2}\,
       \alpha\right)}\over{\sqrt{2}}}
\end{split}
   \ee

\be
   [F_{3,L}]^{1\;2\;1}_{\;1\;2\;1}=[F_{3,L}]^{1\;2\;2}_{\;1\;2\;2}=[F_{3,R}]^{3\;3\;1}_{\;3\;3\;1}=[F_{3,R}]^{1\;3\;1}_{\;1\;3\;1}={{1+q\,\cos \left(\sqrt{2}\,\alpha\right)}\over{\sqrt{2}\,\sqrt{q^2+1}}}
\ee

\be 
 [F_{3,L}]^{1\;3\;1}_{\;1\;3\;1}= [F_{3,L}]^{1\;3\;3}_{\;1\;3\;3}= [F_{3,R}]^{1\;2\;1}_{\;1\;2\;1}=[F_{3,R}]^{2\;2\;1}_{\;2\;2\;1}={{\cos \left(\sqrt{2}\,\alpha\right)+q}\over{\sqrt{2}\,\sqrt{q^2+1}}}
   \ee
      
\be
   [F_{3,L}]^{1\;2\;1}_{\;2\;1\;1}=[F_{3,L}]^{1\;2\;2}_{\;2\;1\;2}=[F_{3,R}]^{1\;1\;2}_{\;1\;2\;1}=[F_{3,R}]^{2\;1\;2}_{\;2\;2\;1}={{\cos \left(\sqrt{2}\,\alpha\right)-q}\over{\sqrt{2}\,\sqrt{q^2+1}}}
\ee

\be
   [F_{3,L}]^{1\;3\;1}_{\;3\;1\;1}=[F_{3,L}]^{1\;3\;3}_{\;3\;1\;3}=[F_{3,R}]^{1\;1\;3}_{\;1\;3\;1}=[F_{3,R}]^{3\;1\;3}_{\;3\;3\;1}={{-1+q\cos \left(\sqrt{2}\,\alpha\right)}\over{\sqrt{2}\,\sqrt{q^2+1}}}
   \ee

    \be
   [F_{3,L}]^{1\;2\;3}_{\;1\;2\;3}=[F_{3,L}]^{1\;3\;2}_{\;1\;3\;2}=[F_{3,R}]^{3\;2\;1}_{\;3\;2\;1}=[F_{3,R}]^{2\;3\;1}_{\;2\;3\;1}={{\left(q+1\right)\,\left(1+2\,\cos \left(\sqrt{3}\,\alpha\right)\right)}\over{3\,
 \sqrt{2}\,\sqrt{q^2+1}}}
\ee

\be
   [F_{3,L}]^{1\;3\;2}_{\;3\;1\;2}=-[F_{3,L}]^{1\;2\;3}_{\;2\;1\;3}=[F_{3,R}]^{2\;1\;3}_{\;2\;3\;1}=-[F_{3,R}]^{3\;1\;2}_{\;3\;2\;1}={{\left(q-1\right)\,\left(1+2\,\cos \left(\sqrt{3}\,\alpha\right)\right)}\over{3\,
 \sqrt{2}\,\sqrt{q^2+1}}}
\ee

\be
\begin{split}
  [F_{3,L}]^{1\;2\;3}_{\;1\;3\;2}&=-[F_{3,L}]^{1\;3\;2}_{\;1\;2\;3}=-[F_{3,R}]^{3\;2\;1}_{\;2\;3\;1}=[F_{3,R}]^{2\;3\;1}_{\;3\;2\;1}=\\
  &{{1-q+\left(q-1\right)\,\cos \left(\sqrt{3}\,\alpha\right)-\sqrt{3}\,\left(1+q
 \right)\,\sin \left(\sqrt{3}\,\alpha\right)}\over{3\,\sqrt{2}\,\sqrt{q^2+1}}}
\end{split}
  \ee

\be
\begin{split}
   [F_{3,L}]^{1\;2\;3}_{\;3\;2\;1}&=-[F_{3,L}]^{1\;3\;2}_{\;2\;3\;1}=-[F_{3,R}]^{1\;3\;2}_{\;2\;3\;1}=[F_{3,R}]^{1\;2\;3}_{\;3\;2\;1}=\\
   &{{1-q+\left(q-1\right)\,\cos \left(\sqrt{3}\,\alpha\right)+\sqrt{3}\,\left(1+q
 \right)\,\sin \left(\sqrt{3}\,\alpha\right)}\over{3\,\sqrt{2}\,\sqrt{q^2+1}}}
\end{split}
   \ee

\be
\begin{split}
  [F_{3,L}]^{1\;2\;3}_{\;2\;3\;1}&=[F_{3,L}]^{1\;3\;2}_{\;3\;2\;1}=[F_{3,R}]^{2\;1\;3}_{\;3\;2\;1}=[F_{3,R}]^{3\;1\;2}_{\;2\;3\;1}=\\
  &{{1+q-\left(q+1\right)\,\cos \left(\sqrt{3}\,\alpha\right)+\sqrt{3}\,\left(q-1\right)
       \,\sin \left(\sqrt{3}\,\alpha\right)}\over{3\,\sqrt{2}\,\sqrt{q^2+1}}}
\end{split}
  \ee

\be
\begin{split}
[F_{3,L}]^{1\;3\;2}_{\;2\;1\;3}&=[F_{3,L}]^{1\;2\;3}_{\;3\;1\;2}=[F_{3,R}]^{1\;2\;3}_{\;2\;3\;1}=[F_{3,R}]^{1\;3\;2}_{\;3\;2\;1}=\\
& {{1+q-\left(q+1\right)\,\cos \left(\sqrt{3}\,\alpha\right)
    -\sqrt{3}\,\left(q-1\right)\,\sin \left(\sqrt{3}\,
    \alpha\right)}\over{3\,\sqrt{2}\,\sqrt{q^2+1}}}
\end{split}
\ee

\normalsize

\section{More on the pentagon identity} \label{pentagonapp}

In this appendix we provide some additional details on the pentagon identity (\ref{pentagon}) for the 3-cocycle $\Phi$
and show how it is guaranteed by the twisting procedure. Let us start by rewriting (\ref{pentagon}) as
\begin{equation}\label{pentagon1}
    \left( \id \otimes \Delta_{F} \otimes \id \right)\Phi = (\mathrm{1} \otimes \Phi^{-1})\cdot \left[\left(\id \otimes \id \otimes \Delta_{F} \right)\Phi \right] \cdot \left[\left( \Delta_{F} \otimes \id \otimes \id \right)\Phi \right]\cdot \left(\Phi^{-1} \otimes \mathrm{1}\right)\;,
\end{equation}
which we note involves the twisted coproduct. By construction the co-associator is given by 
\begin{equation}\label{coassoc}
\Phi = F_{23}\cdot \left[ \left(\id \otimes \Delta \right)(F)\right]\cdot \left[\left(\Delta \otimes \id \right)(F^{-1})\right]\cdot F_{12}^{-1}\;,
\end{equation}
an expression involving the untwisted coproduct. We will use Sweedler notation to write the co-associator as a
sum of tensor products, $\Phi = \sum \phi^{(1)} \otimes \phi^{(2)} \otimes \phi^{(3)}$. We compute: 
\begin{align}\label{coprodcoassc}
\left(\id \otimes \Delta_F \otimes \id \right)\Phi =&\sum \phi^{(1)} \otimes \Delta_{F} (\phi^{(2)}) \otimes \phi^{(3)} \nonumber \\ 
=& \sum \phi^{(1)} \otimes F\Delta (\phi^{(2)})F^{-1} \otimes \phi^{(3)} \nonumber \\
=& \sum\sum \phi^{(1)} \otimes F\left( \phi^{(2)(1)} \otimes \phi^{(2)(2)} \right) F^{-1} \otimes \phi^{(3)} \nonumber \\
=& (\id \otimes F \otimes \id)\left[\sum \phi^{(1)} \otimes \Delta( \phi^{(2)} ) \otimes \phi^{(3)}\right]\cdot\left(\id \otimes F^{-1} \otimes \id \right) \nonumber\\
=& F_{23} \left[\sum \phi^{(1)} \otimes \Delta( \phi^{(2)} ) \otimes \phi^{(3)}\right] F_{23}^{-1} \nonumber\\
=& F_{23} \left[(\id \otimes \Delta \otimes \id)\trr\sum \phi^{(1)} \otimes \phi^{(2)} \otimes \phi^{(3)}\right] F_{23}^{-1} \nonumber\\
=& F_{23} \cdot \left[ \left( \id \otimes \Delta \otimes \id\right) \Phi \right] \cdot F_{23}^{-1}\;.
\end{align}
By the exact same argument one can show that 
\begin{equation}\label{pentagon131}
    \left( \Delta_{F} \otimes \id \otimes \id \right)\Phi \ = \ F_{12} \cdot \left[ \left( \Delta \otimes \id \otimes \id \right)\Phi \right] \cdot F_{12}^{-1}\;, \end{equation}
    \begin{equation}\label{pentagon132} \left(\id \otimes \id \otimes \Delta_{F} \right)\Phi \ = \  F_{34} \cdot \left[\left(\id \otimes \id \otimes \Delta \right)\Phi\right] \cdot F_{34}^{-1} \;.\end{equation}
Let us now express the twist in exponential form:
    \begin{equation}
    F_{qh} =\exp(if_{qh})\;.
\end{equation}
We also define
\begin{align}
F_{qh_{12}} &= F_{qh} \otimes \mathrm{1} = \exp(if_{qh} \otimes \mathrm{1}) = \exp(if_{qh_{12}}) \;, \\
F_{qh_{23}} &= \mathrm{1} \otimes F_{qh} = \exp(i\mathrm{1} \otimes f_{qh} ) = \exp(if_{qh_{23}}) \;.
\end{align}
where the $f_{ij}  \in  \Acal^{\otimes n}$ are defined as $f_{ij} = \mathrm{1} \otimes \dots  \otimes f_{i} \otimes \dots \otimes f_{j} \otimes \dots \otimes \mathrm{1} $. Hence the $f$'s are a tensor chain with the identity element at every site but the $i$-$th$ and $j$-$th$ positions. Suppressing the $(q,h)$ label and absorbing (for clarity)
the imaginary $i$ in the definition of $f$, the action of the coproduct on the twist becomes 
\begin{align}
    \left(\Delta \otimes \id \right)F = \exp\left( f_{13}+f_{23}\right) \ \ \text{and} \ \ \left(\id \otimes \Delta \right)F = \exp(f_{13} + f_{12})\;,
\end{align}
so that the co-associator \eqref{coassoc} and its inverse are given by
\begin{align}
  \Phi = e^{f_{23}} \cdot e^{(f_{13}+f_{12})} \cdot e^{-(f_{13}+f_{23})} \cdot e^{-f_{12}} \ \ \text{and} \ \ \Phi^{-1} = e^{f_{12}} \cdot e^{(f_{13}+f_{23})} \cdot e^{-(f_{13}+f_{12})} \cdot e^{-f_{23}}\;,
\end{align}
from which we obtain 
\begin{equation}\label{eq}
    \Phi^{-1} \otimes \mathrm{1} = e^{f_{12}} \cdot e^{(f_{13}+f_{23})} \cdot e^{-(f_{12}+f_{13})} \cdot e^{-f_{23}}  \ \text{and} \ \mathrm{1} \otimes \Phi^{-1} = e^{f_{23}} \cdot e^{(f_{24}+f_{34})} \cdot e^{-(f_{23}+f_{24})} \cdot e^{-f_{34}}.
\end{equation}
By simply expanding \eqref{pentagon131} and \eqref{pentagon132} we obtain
\begin{align}\label{pentagoncancel}
    \left(\id \otimes \id \otimes \Delta_{F} \right)\Phi \ =& \  F_{34} \cdot \left[\left(\id \otimes \id \otimes \Delta \right)\Phi\right] \cdot F_{34}^{-1}  \nonumber\\
            =& e^{f_{34}} \cdot \left[ e^{(f_{23}+f_{24})} \cdot e^{(f_{12}+f_{13}+f_{14})} \cdot e^{-(f_{13}+f_{14}+f_{23}+f_{24})}\cdot e^{-f_{12}}\right] \cdot e^{-f_{34}} \;,\\
  \left( \Delta_{F} \otimes \id \otimes \id \right)\Phi \ =& \ F_{12}\cdot \left[ \left( \Delta \otimes \id \otimes \id \right)\Phi \right] \cdot F_{12}^{-1} \nonumber \\
            =& e^{f_{12}} \cdot \left[ e^{f_{34}} \cdot e^{(f_{13}+f_{23}+f_{14}+f_{24})} \cdot e^{-(f_{14}+f_{24}+f_{34})} \cdot e^{-(f_{13}+f_{23})} \right] \cdot e^{-f_{12}}\;,
\end{align}
where in the second expression we note that we first had to relabel spaces 2,3 to 3,4 to leave room for the action of $\Delta$ on space 1.
After a few cancellations (noting that of course $e^{-f_{12}}$ commutes with $e^{-f_{34}}$ etc.) the right-hand side of
\eqref{pentagon1} reduces to 
\be
\begin{split}
  \left( \mathrm{1} \otimes \Phi^{-1} \right) &\cdot \left[\left(\id \otimes \id \otimes \Delta_{F} \right)\Phi\right] \cdot \left[\left( \Delta_{F} \otimes \id \otimes \id \right)\Phi \right] \cdot \left( \Phi^{-1} \otimes \mathrm{1} \right)\\
  & = e^{f_{23}}\cdot e^{(f_{24}+f_{34})} \cdot e^{(f_{12}+f_{13}+f_{14})} \times e^{-(f_{14}+f_{24}+f_{34})} \cdot e^{-(f_{12}+f_{13})} \cdot e^{-f_{23}}\;,
\end{split}
\ee
which equals the outcome of expanding the left hand side.
Of course, the fact that the pentagon identity follows from the twisting procedure is known from \cite{Drinfeld90},
but we find explicit derivations such as the one above useful in acquiring confidence with computations in the
quasi-Hopf setting, which can at first appear unfamiliar.

\section{The $G$-tensor} \label{appGtensor}

In this appendix we provide some details on the construction of the $G$-tensor which arises in the mixed star-product
relations of section \ref{MixedRelations}.

We find $G^{i\;j}_{\;k\;l}$ by requiring (\ref{RtGF}), which via a suitable ansatz can easily
be solved for $\widetilde{G}$. Taking the second inverse leads to the tensor $G$. 
It turns out to be simplest to express $G$ in terms of the coefficients $a,\ldots,j$
whose explicit values are given in (\ref{aparam})-(\ref{jparam}). As before, we
will only exhibit the nonzero components where the first index is 1, as cyclically
shifting all indices does not affect their values. We have:

\be
G^{1\;1}_{\;1\;1}=\frac{a~(1-h\hb-h\hb q\qb+q^2\qb^2)-c~(q\qb+h\hb-h^2\hb^2-q^2\qb^2)+d~(1-q\qb+h^2\hb^2-h\hb q\qb)}{2 \left(1-h\hb-q\qb+h^2\hb^2-h\hb q \qb+q^2\qb^2\right)},
\ee
\be
G^{1\;2}_{\;2\;1}=\frac{c~(1-h\hb-h\hb q\qb+q^2\qb^2)-d~(q\qb+h\hb-h^2\hb^2-q^2\qb^2)+a~(1-q\qb+h^2\hb^2-h\hb q\qb)}{2 \left(1-h\hb-q\qb+h^2\hb^2-h\hb q \qb+q^2\qb^2\right)},
\ee
\be
G^{1\;3}_{\;3\;1}=\frac{d~(1-h\hb-h\hb q\qb+q^2\qb^2)-a~(q\qb+h\hb-h^2\hb^2-q^2\qb^2)+c~(1+h^2 \hb^2-h \hb q \qb-q \qb)}{2 \left(1-h\hb-q\qb+h^2\hb^2-h\hb q \qb+q^2\qb^2\right)},
\ee
\be
G^{1\;2}_{\;1\;2}=\frac{(h \hb+q \qb+1) \left(b\,( h \hb \qb+q^2)+f\,\left(\hb^2-h q \qb\right)+j\,(h^2\qb^2+\hb q)\right)}{2(h^3 \qb^3+3 h \hb q \qb-\hb^3+q^3)}\;,
\ee
\be
G^{1\;3}_{\;2\;2}=\frac{(h \hb+q \qb+1) \left(b\, \left(\hb^2-h q \qb\right)+f\,(h^2 \qb^2+\hb q)+j\,(h \hb\qb+ q^2)\right)}{2 \left(h^3 \qb^3+3 h \hb q \qb-\hb^3+q^3\right)}\;,
\ee
\be
G^{1\;1}_{\;3\;2}=\frac{(h \hb+q \qb+1) \left(b\,(\hb q+h^2\qb^2)+ f\,(q^2+ h\hb \qb)+j\,(\hb^2-h q\qb) \right)}{2 \left(h^3 \qb^3+3 h \hb q \qb-\hb^3+q^3\right)}\;,
\ee
\be
G^{1\;3}_{\;1\;3}=-\frac{(h \hb+q \qb+1) \left(e\, (h \qb+ \hb^2 q^2) +g\, \left(h^2-\hb q \qb\right)+i\, (h \hb q+ \qb^2) \right)}{2 \left(h^3-3 h \hb q \qb-\hb^3 q^3-\qb^3\right)}\;,
\ee
\be
G^{1\;1}_{\;2\;3}=-\frac{(h \hb+q \qb+1) \left(e\,(h \hb q+\qb^2)+g\, (h \qb+ \hb^2 q^2)+i\,(h^2 -\hb  q \qb)\right)}{2 \left(h^3-3 h \hb q \qb-\hb^3 q^3-\qb^3\right)}\;
\ee
\be
G^{1\;2}_{\;3\;3}=-\frac{(h \hb+q \qb+1) \left(e \left(h^2-\hb q \qb\right)+g\,( h \hb q+\qb^2)+i\,(h  \qb+\hb^2  q^2)\right)}{2 \left(h^3-3 h \hb q \qb-\hb^3 q^3-\qb^3\right)}\;.
\ee
For star product computations one also requires the second inverse of $G$, which as usual is defined through
$\tilde{G}^{l\;i}_{\;n\;k} G^{j\;k}_{\;l\;m}=\delta^i_{\;m}\delta^{j}_{\;n}$. 

It is certainly possible that the above expressions for $G$ can be further optimised, or that  $G^{i\;j}_{\;k\;l}$ can
be related to $F^{i\;j}_{\;k\;l}$ in a more direct way. As the above expressions are sufficient to demonstrate
the invariance of the K\"ahler part of the action under the $(q,h)$-deformation, we leave further study of the
$G$ tensor for future work.

\bibliography{quasihopf}
\bibliographystyle{utphys}

\end{document}